\documentclass[a4paper,superscriptaddress,showkeys,prX,showpacs,twocolumn]{revtex4-1}
\usepackage{amssymb, amsmath, amsfonts,bm}
\usepackage{graphics,xcolor}
\usepackage{pst-all}
\usepackage{hyperref}
\usepackage[mathlines,running,modulo]{lineno}
\usepackage{relsize}
\usepackage[utf8]{inputenc}
\usepackage{IEEEtrantools}
\usepackage{epsfig}
\usepackage[toc]{appendix}

\begin{document}
%\unitlength = 1mm
%\linenumbers
\title{Theoretical analysis of FMR-driven spin pumping current and its properties via the self-consistent harmonic approximation}
\author{A. R. Moura}
\email{antoniormoura@ufv.br}
\affiliation{Departamento de F\'{i}sica, Universidade Federal de Vi\c{c}osa, 36570-900, Vi\c{c}osa, Minas Gerais, Brazil}
\date{\today}
\begin{abstract}
We applied the self-consistent harmonic approximation (SCHA), combined with coherent states formalism, to  study
the ferromagnetic resonance (FMR) in a ferromagentic/normal metal junction. Due to the interface interaction, 
the FMR-generated spin current is injected from the magnetic insulator to the normal metal, the so-called spin 
pumping. Ordinarily, ferromagnetic models are described by bosonic representation or
phenomenological theories; however, in a coherent magnetization state, the SCHA is the more natural choice to treat FMR
problems. Over the years, the SCHA has successfully applied to investigate ferro- and antiferromagnetism in a wide 
range of scenarios. The main point of the SCHA formalism involves the adoption of a quadratic model for which corrections 
are included through temperature-dependent renormalization parameters. 
Therefore the SCHA is an efficient method for determining the properties of magnetically ordered phases. 
Using the SCHA, we obtained the temperature dependence of FMR-driven spin pumping. 
In addition, we found the spin-mix conductance, the additional damping from the angular momentum injection into 
the normal metal side, and the magnetic susceptibility. The SCHA outcomes are in remarkable 
agreement with the results of the literature.
\end{abstract}
  
\pacs{}
\keywords{Magnetism; spin current; Ferromagnetic Resonance; self-consistent harmonic approximation}

\maketitle

\section{Introduction and motivation}
\label{sec.intro}
The manipulation of spin currents is crucial in spintronic research and has been a topic of great interest
due to its potential application in new spin-based technologies \cite{science294.1488,rmp76.323}. By definition, spin current involves 
an effective transport of angular momentum and, opposite to the conventional (electrical) charge current, the spin current can also
be achieved in insulating materials. A spin current is obtained in conductors when up- and down-oriented electron
spin fluxes show different densities, as occurs in the spin Hall effect \cite{prl83.1834} or using spin valves for filtering
one of the spin-oriented conduction electrons, for example. On the other hand, in magnetic insulators, the spin transport is provided by 
magnons (the quanta of spin wave) \cite{rmp76.323, rezende} or even spinons (neutral half-integer spin 
excitation) \cite{naturephys13.30,prb97.245124,jap123.123903}. Since the current in insulators does not involve charge transport,
it is defined as a pure spin current.

When considering the interface between a normal metal (NM) and a magnetic material (here considered 
a ferromagnetic insulator, FMI) in a junction, two processes deserve special attention. 
The first one is related to the spin current injection from the conductor to the magnetic side due to the spin accumulation
on the normal metal close to the interface, termed spin-transfer torque (SST) \cite{jmmm159.l1,prb54.9353}. Then, in this case, the angular 
momentum injection can induce the magnetization to precess around the ordered axis or even revert its orientation. 
On the other hand, the opposite process, named spin 
pumping (SP), involves the injection of pure spin current from the magnetic side to the conductor \cite{prl88.117601}. The SP process can be 
provided by ferromagnetic resonance (FMR) or electron paramagnetic resonance (EPR), depending on the magnetic sample \cite{prl113.266602}.
In both processes, a resonant magnetic field induces the magnetization to precess and emit angular momentum that is propagated via spin
waves. In addition, due to the magnon absorption, conduction electrons close to the interface are scattered through a spin-flip process.
The EPR/FMR-driven spin pumping is frequently detected on the conductor side by using a metal with strong spin-orbit coupling. Therefore,
due to the inverse spin Hall effect (ISHE) \cite{jap97.10c715,apl88.182509}, the spin current injected is converted into a charge current that
provides a d.c. voltage on the metal. A detailed explanation of SST and SP processes can be found in Refs. \cite{spincurrent12.87,rezende}.

Usually, in addition to the phenomenological description through the Landau-Lifschitz-Gilbert (LLG) equation
\cite{ieee6.3443,prl87.217204,prb66.224403,prl88.117601,prl101.037207,prl111.097602}, 
STT and SP processes have been investigated by adopting Green functions and bosonic representations to describe the magnetic
material \cite{jpcs200.062030,prb89.174417,prb93.064421,prb102.024412}. 
The Holstein-Primakoff representation \cite{pr58.1098,auerbach} allows representing the 
spin operators as first-order creation/annihilation operators only in the magnon low-occupation limit. On the order hand, 
if magnon interactions are relevant, higher-order terms should be considered, 
which introduce complications in the development. In the SU(N) Schwinger 
\cite{prb38.316,prb40.5028,auerbach} representation, each spin component is represented by pair combinations involving N flavors of
bosonic operators, which results in a four-order Hamiltonian. The quartic-order Hamiltonian terms are then decoupled by
introducing auxiliary fields (a mean-field approximation) whose values are determined by solving coupled self-consistent equations.
The Schwinger formalism provides good results in both ordered and disordered phases; however, in frustrated models, including
Gaussian fluctuations becomes necessary \cite{prl78.2216,prb96.174423,prb98.184403,prb100.104431}. In addition, extra care
is required, mainly for three-dimensional models close to the transition temperature \cite{prb66.014407}. The self-consistent
Gaussian approximation (SCGA) \cite{prb53.11593} presents a purpose similar to the self-consistent harmonic approximation (SCHA). 
In the SCGA, the thermodynamics of a classical spin model is evaluated through self-consistent equations
depending on the magnetization and their quadratic fluctuations. In this case, the Gaussian corrections are introduced by 
considering spin cumulants \cite{pr124.1757,pr130.155} in the statistical averages. The SCGA formalism provides good results; 
however, the number of self-consistent parameters is larger than the SCHA, and the quantization is more challenging to implement.
 On the other hand, the SCHA provides a simple quadratic method in which corrections are implemented by renormalization parameters
depending on temperature. The renormalization parameters are self-consistently determined in order to give the best harmonic
approximation in terms of $S^z$ spin component and its conjugated angle $\varphi$. More details about the SCHA formalism
are given in Sec. (\ref{sec.scha}).

\begin{figure}[h]
\centering \epsfig{file=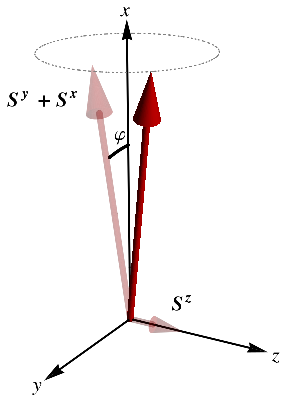,width=0.5\linewidth}
\caption{The magnetization is aligned along the $x$ axis (the direction of a static magnetic field $B^x$) and
$S^y, S^z\ll S^x$. Here, $\varphi$ is defined as the angle between the spin projection on the $xy$ plane and the $x$ axis.}
\label{fig.single_spin}
\end{figure}

In FMR experiments, the magnetization precession exhibits a coherent phase of the spin field just like the electromagnetic 
field does in a LASER. In this case, the entire spin field shows synchronous dynamics and can be represented by using a 
single spin that is well pictured by a classical vector. A quantum state like that is formally described by a coherent state, 
which was initially used to derive a fully quantum model of the radiation fields \cite{pr131.2766,gerry} as well 
as the coherent behavior of magnons \cite{pla29.47,pla29.616,prb4.201}. It is
well-known that coherent states represent the more classical-quantum state, {\it i.e.}, states with minimum uncertainty \cite{rmp62.867}. 
Consider, for example, a particle in a harmonic potential and represented by a coherent state. In this case, $\Delta x\Delta p=\hbar/2$, 
while the wave function describes a dispersionless wave packet that moves harmonically around the minimum of the potential. 
Similar semiclassical behavior is reached for the spin field in the FMR. Then, we represent the spin by the phase 
angle $\varphi$ around the $z$ axis and the conjugate momentum associated, namely $S^z$. In some cases, $S^z$ is aligned 
with the magnetization direction, but this is unnecessary. Here, we define $S^z$ and $S^y$ as transverse 
components throughout the text, while the magnetization is along the $x$ axis as shown in Fig. \ref{fig.single_spin}. 
Note that $\varphi\ll 1$ and thus, the transverse spin components $S^y$ and $S^z$ are much smaller that the longitudinal 
component $S^x$. In addition, provided that $S^z\propto \dot{\varphi}$, both fields $\varphi$ and $S^z$ show an oscillating behavior 
during the magnetization precession as it is explained in next section. 
From the classical point of view, the fields $\varphi_i$ and $S^z_j$, on sites $i$ and $j$, respectively, 
satisfy the Poisson bracket $\{\varphi_i,S_j^z\}=\delta_{ij}$, and the quantization is achieved by promoting the fields to operators 
that obey the commutation relation $[\varphi_i,S_j^z]=\delta_{ij}$. Similar to the particle case, the operators obey the local
equality $\Delta\varphi\Delta S^z=1/2$, which justifies the semiclassical magnetization behavior of the spin. Therefore it is natural 
to adopt $\varphi$ and $S^z$ as the fundamental operators for describing magnetic models in FMR experiments instead of the 
usual bosonic representations.

Over the years, the self-consistent harmonic approximation has been successfully applied to evaluate the critical 
temperature \cite{prb49.9663,pla202.309,prb51.16413,prb54.3019,ssc104.771,prb59.6229}, 
the topological BKT transition \cite{pla166.330,prb48.12698,prb49.9663,prb50.9592,ssc100.791,prb53.235,prb54.3019,prb54.6081,ssc112.705,epjb2.169,pssb.242.2138,prb78.212408,jmmm452.315}, 
and the large-D quantum phase transition \cite{pasma373.387,jpcm20.015208,pasma388.21,pasma388.3779,jmmm357.45} in a wide variety 
of magnetic models. In the SCHA formalism, the Hamiltonian is expanded to second order in $\varphi$ and $S^z$ operators, 
while higher-order contributions are included through temperature-dependent renormalization parameters. 
In addition, Moura and Lopes have demonstrated that SCHA is entirely compatible with the coherent state approach \cite{jmmm472.1}. 
Therefore the SCHA formalism is the most plausible choice for studying the magnetization precession phenomena. 
In this work, we used the SCHA formalism to provide a new framework for the FMR-driven spin pumping across an NM/FMI junction interface. 
As primary outcomes, we obtain the FMR-driven spin current across the interface, the spin-mixing conductance, the additional
Gilbert damping due to the angular momentum injection, and the magnetic susceptibility. All our results are in excellent 
agreement with well-known results in the literature.

\section{Model description}
\label{sec.model}
\begin{figure}[h]
\centering \epsfig{file=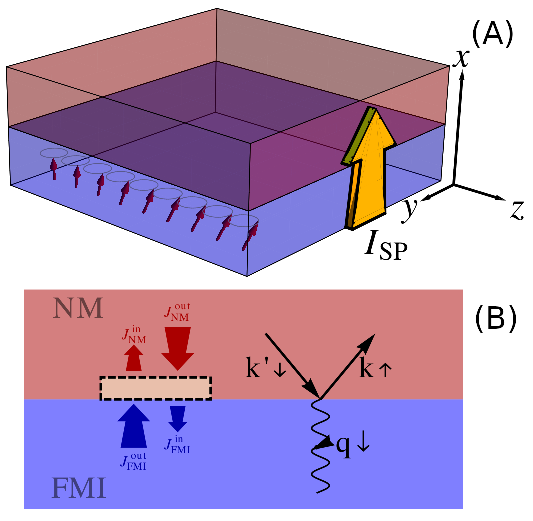,width=0.9\linewidth}
\caption{(a) The NM/FMI junction and the adopted orientation of the axis. (b) The spin current across a pillbox at the FMI/NM interface, 
and the diagram representing the interface representation.}
\label{fig.interface}
\end{figure}

In the present work, we consider a NM/FMI junction. The ferromagnetic material is a thin film with the
magnetization axis (defined as the $x$ axis) normal to the film plane, as shown in Fig. \ref{fig.interface}.
After minor modifications, the case whose magnetization is parallel to the plane could also be investigated through 
the SCHA formalism. The electronic side is considered a nonmagnetic spin sink, as platinum. For a poor spin sink, the conduction-electron 
spin-diffusion length is large, and a spin accumulation takes place close to the interface, which results in a spin black-flow 
into the FMI \cite{prb66.224403,rmp77.1375}. However, we are mainly concerned with the spin pumping process, which we consider a 
perfect spin sink, and conduction electrons rapidly decay after spin-flip scattering at the interface. Therefore there is 
no relevant spin accumulation, and the spin back-flow can be disregarded. In addition, bulk electronic interactions are also 
supposed to be unessential, and a free electron model represents the normal metal.

Conduction electrons interact with localized electrons at the interface through an sd-exchange 
potential \cite{jpcs200.062030,mahan}. Thus the Hamiltonian is written as the sum
$H=H^m+H^e+H^{sd}$, where $H^m$, $H^e$, and $H^{sd}$ are the magnetic, 
electronic, and interface contributions, respectively. The usual ferromagnetic Heisenberg model gives the magnetic Hamiltonian
\begin{equation}
\label{eq.Hm}
H^m=-\frac{J}{2}\sum_{\langle ij\rangle}{\bf S}_i\cdot{\bf S}_j-g\mu_B\sum_i {\bf S}_i\cdot{\bf B}_i(t)
\end{equation}
where $J>0$ is the exchange coupling, and the first sum is done over nearest neighbors. 
${\bf B}_i(t)=\mu_0(H_i^x-N_x M_x)\hat{\textrm{\it \i}}+\mu_0 H_i^y(t)\hat{\textrm{\it \j}}+\mu_0 H_i^z(t)\hat{k}$ is the effective
magnetic field, which is composed of the external field ${\bf H}$ and the demagnetizing field oriented along the $x$ axis. 
$M_x$ is the normal magnetization and, due to the adopted geometry, $N_x=1$, while $N_y=N_z=0$. 
In the above equation, $H_i^x$ is a constant field responsible for aligning the spin field 
while the transverse components, $H_i^{y,z}(t)$, are oscillating fields that induce the magnetization precession. 
Here, we have included only the terms necessary to reach the coherent behavior; however, 
other contributions, such as different anisotropies, can be implemented to improve the model.
As will be justified in the next section, the axis was chosen to provide a simpler development in the SCHA formalism.
In the many-body representation, the electronic Hamiltonian is expressed as
\begin{equation}
H^e=\sum_{k\sigma}\epsilon_k c_{k\sigma}^\dagger c_{k\sigma},
\end{equation}
where $\epsilon_k=\hbar^2k^2/2m$, $c_{k\sigma}$ ($c_{k\sigma}^\dagger$) is the annihilation (creation) electron operator, and
$\sigma=\uparrow,\downarrow$ is the spin index. Here, as usual, we adopt electron spins aligned along the magnetization direction ($x$ axis). 
Therefore the electron states are defined as the eigenstates of $\sigma_x$. Finally, the interface interaction is given by
\begin{equation}
H^{sd}=2J_{sd}\sum_i {\bf s}_i\cdot{\bf S}_i,
\end{equation}
where $J_{sd}<0$ is the coupling between conduction and localized electrons, $s_i$ is the conduction electron spin operator,
$S_i$ is the spin of localized electrons on the FMI, and the sum is done over
the interface sites. As we will see in the following sections, the injected spin current is highly dependent on the 
sample properties, including the interface coupling. Since sd-exchange is sensitive to the electron distance interaction, it
is not easy to stipulate an exact value over the entire surface. An estimated value for the sd coupling is of 
the order of $-0.1 eV$ \cite{ptp32.37}. In addition, ${\bf s}_i=\psi_i^\dagger{\bm \sigma}\psi_i$, where
$\psi_i^\dagger=(c_{i\uparrow}^\dagger\ c_{i\downarrow}^\dagger)$ is the electron spinor and ${\bf \sigma}$ is a vector 
whose components are given by the Pauli matrices. Using the basis of $\sigma_x$ eigenstates, in the momentum space, 
$H^{sd}$ is written as
\begin{equation}
H^{sd}=\sum_{kk^\prime q}\Lambda_{kk^\prime q}S_q^+ c_{k^\prime\downarrow}^\dagger c_{k\uparrow}+H.c.
\end{equation}
with $S_q^+=S_q^y+iS_q^z$ and
\begin{equation}
\Lambda_{kk^\prime q}=\frac{J_{sd}}{N_e \sqrt{N_m}}\sum_i e^{i({\bf k}-{\bf k^\prime}-{\bf q})\cdot{\bf r}_i},
\end{equation}
where $N_e$ and $N_m$ are the number of conduction electrons and magnetic sites, respectively. 
In the above equation, we consider only the terms that imply spin-flip scattering (related to the 
spin transverse components). The longitudinal term $s_i^x S_i^x$ involves number particle conservation scatterings 
and does not contribute to the spin current across the interface. Indeed, the injection (or absorption) of
angular momentum is related to a change of the magnetization component along the 
angular momentum direction ($M^x$ in our case). It is possible only for interaction terms that includes the $S^+$ and $S^-$
ladder operators . In addition, the spin-flip scattering is related to the spin-mixing conductance, which arises in the 
LLG formalism and represents the transparency of the spin current across the interface \cite{prb66.224403,prl88.117601}. 

\section{SCHA formalism}
\label{sec.scha}
As commented previously, in the coherent magnetization phase, the more natural spin representation 
is done by using $\varphi$ and $S^z$ as fundamental operators, which is achieved through the Villain representation 
$S_i^+=e^{i\varphi_i}\sqrt{\tilde{S}^2-S_i^z(S_i^z+1)}$, where $\tilde{S}=\sqrt{S(S+1)}$ \cite{jp35.27}. Therefore 
one can always expand the spin components up to second order in $\varphi$ and $S^z$ to provide the spin wave spectrum energy
without any correction. However, better results are obtained with the inclusion of renormalization parameters that 
consider the contributions of higher-order terms. In the SCHA, we include a renormalization factor $\rho$ for each term
that presents a phase expansion. Therefore, in the series expansion, we replace $\varphi$ by $\sqrt{\rho}\varphi$. 
The renormalization parameters are then found by solving a set of self-consistent equations. 
Here, we treat the time-dependent term of the magnetic Hamiltonian as a potential, solved in Sec. \ref{sec.mcs}, 
while the quadratic model represents the constant contribution
\begin{IEEEeqnarray}{rCl}
H_0^m&=&\frac{J}{2}\sum_{\langle ij\rangle}\left(\frac{\rho_{E}\tilde{S}^2}{2}\Delta\varphi^2+S_i^z S_i^z-S_i^z S_j^z\right)+\nonumber\\
&&+\frac{g\mu_B B_x}{2}\sum_i\left(\rho_{B}\tilde{S}\varphi_i\varphi_i+\frac{1}{\tilde{S}}S_i^z S_i^z \right),
\end{IEEEeqnarray}
where we adopt a uniform field $H^x$, $\Delta\varphi=\varphi_j-\varphi_i$, and include one factor renormalization for 
each $\varphi$ expansion. Generally, $\rho_{E}\lesssim\rho_{B}$, and both parameters abruptly vanish at the same critical temperature.
In momentum space, the Hamiltonian assumes the simple quadratic form
\begin{equation}
\label{eq.H0m}
H_0^m=\frac{1}{2}\sum_q\left(h_q^\varphi \rho_{\textrm{eff}} \tilde{S}^2\varphi_{-q}\varphi_q+h_q^z S_{-q}^z S_q^z\right),
\end{equation}
where $\rho_{\textrm{eff}}=\sqrt{\rho_{E}\rho_{B}}$ is an effective renormalization parameter, the coefficients are given by
\begin{IEEEeqnarray}{rCl}
\IEEEyesnumber
\IEEEyessubnumber*
h_q^\varphi&=&z J(1-\gamma_q)\sqrt{\frac{\rho_{E}}{\rho_{B}}}+\frac{g\mu_B B_x}{\tilde{S}}\sqrt{\frac{\rho_{B}}{\rho_{E}}},\\
h_q^z&=&zJ(1-\gamma_q)+\frac{g\mu_B B_x}{\tilde{S}},
\end{IEEEeqnarray}
and $\gamma_q=z^{-1}\sum_{\bm \eta}e^{i{\bf q}\cdot{\bm \eta}}$ is the factor structure of $z$ nearest-neighbor 
spins located at ${\bm \eta}$ positions. Note that, using the ansatz $S_q^y\approx\sqrt{\rho_{\textrm{eff}}}\tilde{S}\varphi_q$, 
we can write the Hamiltonian in terms of fluctuations of the transverse spin components as 
$H_0^m=(1/2)\sum_q(h_q^\varphi S_{-q}^y S_q^y+h_q^z S_{-q}^z S_q^z)$. From the semiclassical analysis,
the spin dynamics is obtained from the Hamilton equations $\dot{\varphi}_{-q}=-\partial H/\partial S_q^z$ and 
$\dot{S}_{-q}^z=\partial H/\partial\varphi_q$, which provide the transverse spin component dynamics
\begin{IEEEeqnarray}{rCl}
\IEEEyesnumber
\IEEEyessubnumber*
\dot{S}_q^y&=&-\sqrt{\rho_\textrm{eff}}\tilde{S}h_q^z S_q^z,\\
\dot{S}_q^z&=&\sqrt{\rho_\textrm{eff}}\tilde{S}h_q^\varphi S_q^y.
\end{IEEEeqnarray}
Assuming the elliptical oscillating transverse behavior, {\it i.e.}, $S_q^y(t)=A_q^y\cos(\omega_q t)$ and 
$S_q^z(t)=A_q^z\sin(\omega_q t)$, where $A_q^{y,z}$ are the transverse amplitudes, it is easy to obtain the spin wave 
frequencies, given by $\omega_q=\tilde{S}\sqrt{\rho_\textrm{eff}h_q^\varphi h_q^z}$.

To diagonalize the quantum Hamiltonian, we define bosonic operators via
\begin{IEEEeqnarray}{rCl}
\label{eq.a}
\IEEEyesnumber
\IEEEyessubnumber*
\varphi_q&=&\frac{1}{\sqrt{2}}\left(\frac{h_q^z}{\rho_{\textrm{eff}}\tilde{S}^2h_q^\varphi}\right)^{1/4}(a_q^\dagger + a_{-q})\\
S_q^z&=&\frac{i}{\sqrt{2}}\left(\frac{\rho_{\textrm{eff}}\tilde{S}^2h_q^\varphi}{h_q^z}\right)^{1/4}(a_q^\dagger - a_{-q}),
\end{IEEEeqnarray}
which results in $H_q^m=\sum_q E_q a_q^\dagger a_q$, where
\begin{equation}
E_q=\hbar\omega_q=\tilde{S}\sqrt{\rho_{\textrm{eff}}h_q^\varphi h_q^z}
\end{equation}
are the magnon eigenenergies, in agreement with the semiclassical result. 
In addition, it is a straightforward procedure to get the Holstein-Primakoff-like
ladder operators
\begin{equation}
\label{eq.Splus}
S_q^+\approx\sqrt{\rho_\textrm{eff}}\tilde{S}\varphi_q+iS_q^z=\sqrt{2\tilde{S}}\rho_{\textrm{eff}}^{1/4} b_q,
\end{equation}
and $S_q^-\approx\sqrt{2\tilde{S}}\rho_\textrm{eff}^{1/4}b_q^\dagger$, where we define
\begin{equation}
b_q=\cosh\theta_q a_q+\sinh\theta_q a_{-q}^\dagger
\end{equation}
with the angle $\theta_q$ is determined from
\begin{equation}
\tanh\theta_q=\frac{\sqrt{h_q^z}-\sqrt{h_q^\varphi}}{\sqrt{h_q^z}+\sqrt{h_q^\varphi}}.
\end{equation}
Note that states generated by the $b$ operator are linear combinations of $a$ states moving in opposite directions.
Since $a_q$ operators diagonalize the transverse spin component Hamiltonian, the $a$ states represent modes
with spin in the $yz$ plane. In contrast, $b$ states are magnons with spin along the $x$ axis. When we disregarded
the renormalization procedure (equivalent to adopt $\rho_E=\rho_B=1$), we obtain $a_q=b_q$, provided that $\theta_q=0$.
Since $h_q^\varphi=h_q^z$, the Hamiltonian can be written only in terms of $S_{-q}^x S_q^x$, and it is natural
to consider spin fluctuation along the longitudinal direction. The renormalization parameters are given by
\begin{equation}
\rho_{E}=\left(1-\frac{\langle S_i^z S_i^z\rangle_0}{\tilde{S}^2}\right)e^{-\frac{1}{2}\langle\Delta\varphi^2\rangle_0},
\end{equation}
and
\begin{equation}
\rho_{B}=\left(1-\frac{\langle S_i^z S_i^z\rangle_0}{2\tilde{S}^2}\right)e^{-\frac{1}{2}\langle\varphi_i\varphi_i\rangle_0},
\end{equation}
where the indexes $E$ and $B$ states for the exchange and the static magnetic field ($B^x$) contribution, respectively. 
A quick demonstration of the above equations is given in Appendix (\ref{appendix1}). In addition, to determine 
$\rho_E$ and $\rho_B$, we must also resolve the equation of the magnetization $M^x=(g\mu_B/a^3)\langle S^x\rangle$, 
where $a^3$ is the unit cell volume and
\begin{equation}
\langle S^x\rangle=\frac{1}{2}\langle(S^++S^-)\rangle_0\approx\tilde{S}\rho_B.
\end{equation}
At finite temperatures, the expected values are determined by the statistical average using Eq. (\ref{eq.a}). At a critical
temperature $T_c$, both parameters abruptly drop to zero, and so the SCHA is suitable only for $T<T_c$. 

\section{Magnetic Coherent state}
\label{sec.mcs}
The static ($H^x$) and dynamic ($H^y$ and $H^z$) components of field ${\bf H}$ are fundamental pieces to provide the FMR-driven spin pumping. 
In a typical FMR experiment, an alternating field at microwave frequencies forces the spin field to oscillate around the 
direction defined by the static field perpendicular to the dynamic one. While the frequency $\Omega$ of the oscillating 
field is kept constant, the static field is adjusted to provide the resonance condition of the excited magnons, {\it i.e.}, 
$\Omega=\omega_{q=0}$. When the resonance condition is achieved, the entire spin field oscillates in the synchronous behavior, 
which defines the coherent magnetization state. In this section, we show that the SCHA provides an efficient formalism to
describe the coherent phase of FMR experiments. 

To adequately describe the role of the oscillating field, we consider the Zeeman energy associated with it as a time-dependent 
potential expressed, in the momentum space, as
\begin{equation}
V(t)=-\frac{g\mu_B\sqrt{2\tilde{S}}}{2}\sum_q\left[ S_q^+ B_q^-(t)+S_q^- B_q^+(t)\right],
\end{equation}
where $B_q^+(t)=B_q^y(t)+i B_q^z(t)\equiv B_q e^{-i\Omega t}$. The time evolution is then written as
$\hat{A}(t)=S^\dagger(t)\hat{A}_0(t)S(t)$, where
\begin{equation}
\hat{A}_0(t)=e^{\frac{i}{\hbar} K_0 t}\hat{A} e^{-\frac{i}{\hbar}K_0 t}
\end{equation}
with $K_0=K_0^e+H_0^m$, and we define the time evolution operator
\begin{equation}
\label{eq.S_V0}
S(t)=T_t \exp\left[-\frac{i}{\hbar}\int_0^t \hat{V}_0(t^\prime)\right]dt^\prime.
\end{equation}
In this case, opposite to the LRT procedure, it is unnecessary to expand the exponential in lower orders of $\hat{V}_0$. Using
the Eq. (\ref{eq.Splus}), the exponential argument is expressed as $\sum_q \left( \bar{\alpha}_q a_q-\alpha_q a^\dagger_q\right)$,
where the coefficient $\alpha_q$ is given by
\begin{IEEEeqnarray}{rCl}
\alpha_q&=&i\gamma\sqrt{2\tilde{S}} \rho_\textrm{eff}^{1/4}B_q \int_0^t\left(\cosh\theta_q e^{-i\Omega t^\prime}\right.+\nonumber\\
&&\left.+\sinh\theta_qe^{i\Omega t^\prime}\right)e^{i(\omega_q+i\varepsilon)t^\prime}dt^\prime,
\end{IEEEeqnarray}
with $\varepsilon\ll\omega_q$ being an infinitesimal factor included to ensure the convergence for long times, and 
$\gamma=g\mu_B/\hbar$. The convergence factor plays the same role as a damping term, which was not considered {\it a priori} but 
can be added through a phenomenological analysis. 

Usually, the NM/FMI samples are tiny, and the oscillating fields can be considered uniform over the magnetic
material, which results in $B_q=\sqrt{N_m}B_\perp\delta_{q,0}$. In addition, since $\Omega$ is of the order of 10 GHz, we can adopt that 
$\Omega^{-1}\ll \varepsilon^{-1}\ll t$, which simplifies the integral result to the time-independent value
$\alpha_q=\alpha_0\delta_{q,0}$, with
\begin{equation}
\alpha_0=\gamma B_\perp\sqrt{2\tilde{S}N_m}\rho_\textrm{eff}^{1/4} \left(\frac{\cosh\theta_0}{\Omega-\omega_0-i\varepsilon}-\frac{\sinh\theta_0}{\Omega+\omega_0+i\varepsilon}\right).
\end{equation}
Therefore, provided the high frequency of the oscillating field, the system rapidly assumes a stationary regime with
uniform magnetization precession. The time evolution operator, given by Eq. (\ref{eq.S_V0}), assumes 
a time-independent limit when $\varepsilon t\gg 1$, and so we write $S(t\gg\varepsilon^{-1})=D(\alpha)$, where
\begin{equation}
D(\alpha)=\exp\left[\sum_q \left( \alpha_q a^\dagger_q-\bar{\alpha}_q a_q\right)\right],
\end{equation}
is the displacement operator that defines a coherent state by $|\alpha\rangle=D(\alpha)|0\rangle$,
with $|0\rangle$ being the vacuum state. At finite temperature, the thermodynamics of coherent states is
given by the thermal coherent states \cite{pla134.273,jmo38.2339}, which asserts that statistical averages
are obtained from $\langle \hat{A}\rangle=\textrm{Tr}(\rho_{cs}\hat{A})$, where $\rho_{cs}=D(\alpha)\rho_0 D^\dagger(\alpha)$
defines the coherent state density matrix, with $\rho_0=e^{-\beta K_0}/\textrm{Tr}\rho_0$.
Therefore, using the property $D^\dagger(\alpha)a_q D(\alpha)=a_q+\alpha_q$, 
we obtain $\langle a_q\rangle=\alpha_q$, and $\langle a^\dagger_q a_q\rangle=n(E_q)+|\alpha_q|^2$,
where $n(E_q)=(e^{\beta E_q}-1)^{-1}$ is the Bose-Einstein distribution, which counts the thermal excited states, and
$|\alpha_q|^2=N_q$ is the number of modes in the condensate state (usually the $q=0$ state). 
It is important to observe that in a coherent phase, a finite fraction of the particles (or excitation modes) occupy the same 
coherent state and $N_q\approx N$. In contrast, other states have a very low occupations, and we can disregard them.

\begin{figure}[h]
\centering \epsfig{file=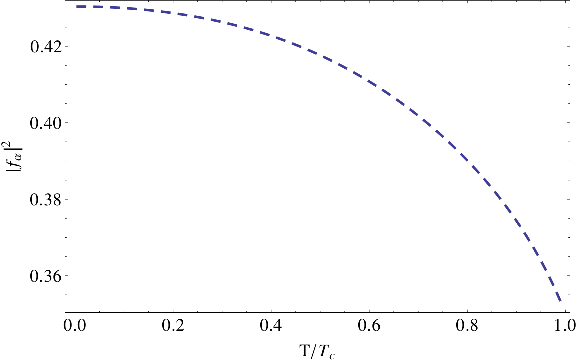,width=0.9\linewidth}
\caption{The coherent occupation level $|f_\alpha|^2$ as function of the temperature. At $T=0$, approximately 43$\%$ of
the magnons are in the condensate state.}
\label{fig.fa}
\end{figure}

Close to the resonance condition, we can use $\Omega\approx\omega_0$, and write $\alpha_0(T)=f_\alpha(T) \sqrt{N_m}$, where
\begin{equation}
f_\alpha(T)=\frac{\sqrt{2\tilde{S}}\gamma B_\perp(\rho_E^{1/4}+\rho_B^{1/4})}{4(\Omega-\omega_0-i\epsilon)},
\end{equation}
measures the occupation of the coherent state. Replacing the infinitesimal parameter $\epsilon$ by $\eta_0\Omega$, where 
we adopt a typical value of the order of $\eta_0\sim 10^{-3}$, we obtain $\gamma B_\perp/\eta\Omega\approx 1$, and so 
$|\alpha_0|^2\approx 0.4 N_m$. Note that a vanishing dissipation parameter implies in nonphysical behavior since the
model acquires infinite energy due to the oscillating field. When the temperature increases, 
the number of magnons in the condensate phase decreases, and at $T=T_c$, the coherent state vanishes. For
$T>T_c$, there is no mode in the condensate state, and magnetic excitations are composite only by thermal magnons (with distribution
following the Bose-Einstein statistics). Figure {\ref{fig.fa} shows the dependence on the temperature of the occupation level.
Curiously, even at zero temperature, only a fraction of the magnons participates in the coherent phase, similar to 
the physics of $^4$He superfluid, for example.

\begin{figure}[h]
\centering \epsfig{file=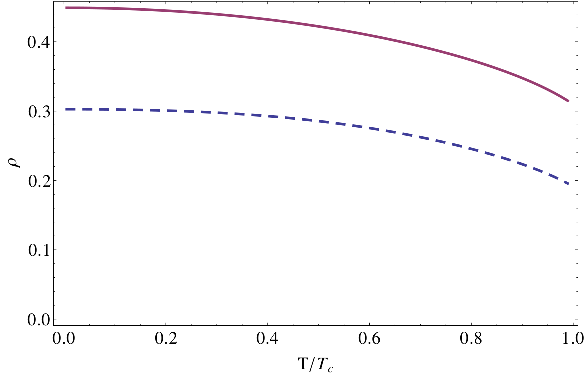,width=0.9\linewidth}
\caption{The plot shows temperature dependence of the renormalization parameters $\rho_E$ 
(dashed line) and $\rho_B$ (solid line). $T_c=1.83 J/k_B$ is the critical temperature 
where the parameters abruptly vanish.}
\label{fig.rho}
\end{figure}

The SCHA correctly describes the oscillating behavior of the transverse spin component when we consider the
coherent state development. Indeed, using Eq. (\ref{eq.a}), we obtain the transverse magnetization dynamics
\begin{IEEEeqnarray}{rCl}
\IEEEyesnumber
\IEEEyessubnumber*
\label{eq.magnetization}
M_q^y(t)&=&A_q^y\cos (\Omega t-\phi_0),\\
M_q^z(t)&=&-A_q^z\sin (\Omega t-\phi_0),
\end{IEEEeqnarray}
where the transverse amplitudes are defined by
$A_q^y=(g\mu_B/a^3)\sqrt{2\tilde{S}}\rho_E^{1/4}|\alpha_0|\delta_{q,0}$,
and $A_q^z=(g\mu_B/a^3)\sqrt{2\tilde{S}}\rho_B^{1/4}|\alpha_0|\delta_{q,0}$, while $\phi_0$ is
the phase of $\alpha_0$. Note that, due to the adopted representation, the magnetization shows
a clockwise rotation, opposite to the usual counter clockwise direction. In addition, the averages 
present in the self-consistent equations are determined using
\begin{equation}
\langle S_{-q}^zS_q^z\rangle_{cs}=\frac{\tilde{S}}{2}\sqrt{\frac{\rho_\textrm{eff}h_q^\varphi}{h_q^z}}\coth\left(\frac{\beta\hbar
\omega_q}{e}\right)+\langle S_q^z\rangle_{cs}^2
\end{equation}
and
\begin{equation}
\langle\varphi_{-q}\varphi_q\rangle_{cs}=\frac{1}{2\tilde{S}}\sqrt{\frac{h_q^z}{\rho_\textrm{eff}h_q^\varphi}}\coth\left(\frac{\beta\hbar\omega_q}{2}\right)+\langle\varphi_q\rangle_{cs}^2,
\end{equation}
where the hyperbolic cotangent term is related to the usual thermal fluctuations, while $\langle S_q^z\rangle_{cs}^2$ 
and $\langle\varphi_q\rangle_{cs}^2$ are finite only in the precession stat and measure the coherent phase. To solve the 
self-consistent equations, we also assume a time average and replace $\cos^2\Omega t$, and $\sin^2\Omega t$ by $1/2$. 
Considering $S=1$, $\mu_0 H^x=0.1 T$, $\mu_0 H_\perp=10^{-4} T$, and lattice spacing $a= 10^{-9} m$, we determine the 
renormalization parameters and its dependence on temperature is shown in Fig. \ref{fig.rho}.
Both parameters abruptly drop to zero at the critical temperature $T_c=1.83 J/k_B$ and, for $T<T_c$, $\rho_E\lesssim\rho_B$. 
The critical temperature was determined considering the bcc lattice with a single ion per unit cell, and 
other configurations provide a different ratio $k_B T_c/J$. Here, we express the energies in units 
of $J$, and typical values of the exchange coupling are between $10^{-5}$ to $10^{-3}$ eV.
Including anisotropic terms or other weak interactions also slightly changes the ratio $k_B T_c/J$. 
However, the results obtained from the simpler Hamiltonian (\ref{eq.Hm}) are in agreement with expected experimental measurements.
We also determine the magnetization $M^x$ dependence on temperature, and Fig. \ref{fig.Mx} shows the result obtained. 
\begin{figure}[h]
\centering \epsfig{file=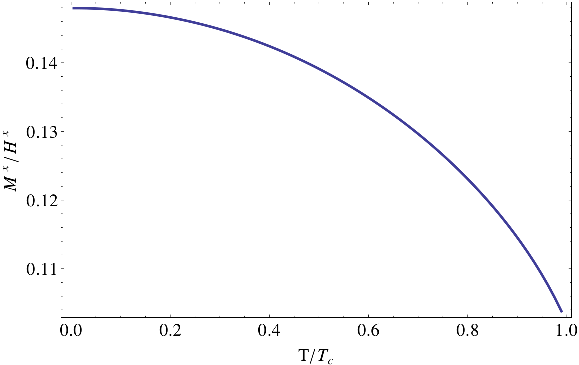,width=0.9\linewidth}
\caption{The magnetization $M^x$ (in units of $H^x$) of the ferromagnetic thin film below the
critical temperature $T_c=1.83 J/k_B$.}
\label{fig.Mx}
\end{figure}

\section{spin current through the interface}
\label{sec.spincurrent}
The spin current across the interface can be evaluated on any side of the NM/FMI junction. Therefore, to 
determine the spin current, we consider a pillbox, on the NM side, in contact with the interface as 
shown in Fig. \ref{fig.interface}. The spin current across the pillbox boundary is composed of in 
and out components of spin current on the NM ($J_\textrm{NM}$) and FMI ($J_\textrm{FMI}$) sides. 
For a pillbox with a height much smaller than the conduction-electron spin-diffusion length, we can 
disregard bulk spin-flip scattering, and the continuity equation provides 
$I_s=I_\textrm{STT}-I_\textrm{SP}=-(\hbar/2)\partial_t(N_\uparrow^e-N_\downarrow^e)$, where we define 
$I_\textrm{STT}=I_\textrm{FMI}^\textrm{(in)}-I_\textrm{NM}^\textrm{(out)}$ and $I_\textrm{SP}=I_\textrm{FMI}^\textrm{(out)}-I_\textrm{NM}^\textrm{(in)}$.
Eventually, we will adopt conditions that vanish the spin current from the NM side, the STT contribution, 
and consider only the FMR-driven spin current. Thus, using the Heisenberg equation of motion, we obtain the spin 
current operator $I_s=i(A-A^\dagger)$, where
\begin{equation}
A^\dagger=\sum_{kk^\prime q}\Lambda_{kk^\prime q}S_q^+ c_{k^\prime\downarrow}^\dagger c_{k\uparrow}.
\end{equation}
The expected value of the spin current is determined in the interaction picture by 
$\langle I_s(t)\rangle=\langle S^\dagger(t) \hat{I}_s(t) S(t)\rangle$, where the caret stands for time evolution according 
to the Hamiltonian $H-H^{sd}$, while, for small coupling at the interface, the time evolution operator $S(t)$ is approximated by
\begin{equation}
S(t)\approx 1-\frac{i}{\hbar}\int_{-\infty}^t \hat{H}^{sd}(t^\prime)dt^\prime,
\end{equation}
where we adopt an adiabatic evolution from $t\to-\infty$ (when $H^{sd}=0$) to $t=0$. Therefore the spin current is given by
\begin{equation}
I_s=\frac{2}{\hbar}\textrm{Im}\int_{-\infty}^{\infty}i\theta(t)\langle[\hat{A}(t),\hat{A}^\dagger(0)]\rangle dt.
\end{equation}
Note that electronic states have time evolution according to $H^e$. At the same time, the statistical average, as usual, are 
evaluated using the grand canonical Hamiltonian $K^e=\sum_{k\sigma}\xi_{k\sigma}c_{k\sigma}^\dagger c_{k\sigma}$, where
$\xi_{k\sigma}=\epsilon_k-\mu_\sigma$, and $\mu_\sigma$ is the chemical potential for electrons with spin $\sigma$. It is
convenient to replace the time evolution to match with the Boltzmann weight, which provides 
\begin{equation}
\label{eq.Is}
I_s=-2\textrm{Im}U_\textrm{ret}\left(\delta\mu\right),
\end{equation}
where $\delta\mu=\mu_\uparrow-\mu_\downarrow$, and $U_\textrm{ret}(\delta\mu)$ is the time Fourier transform
\begin{equation}
U_\textrm{ret}\left(\delta\mu\right)=\int_{-\infty}^{\infty} U_\textrm{ret}(t) e^{\frac{i}{\hbar}\delta\mu t}dt
\end{equation}
of the retarded function
\begin{equation}
\label{eq.Uret}
U_\textrm{ret}(t)=-\frac{i}{\hbar}\langle[\hat{A}(t),\hat{A}^\dagger(0)]\rangle.
\end{equation} 
In the above equation, despite the same notation, the time evolution is defined by using $K_e$, while
the electron energies are measured in relation to the chemical potential $\mu_\sigma$. 
In this work, as we are interested in the FMR-drive spin current, from now on,
we consider a perfect spin sink, which implies $\delta\mu=0$, and consequently $I_\textrm{STT}=0$.
It is easy to obtain the retarded Green's function, whose magnetic part is now evaluated by using the coherent states
obtained from the previous section. Using $\hat{A}=D^\dagger(\alpha)\hat{A}_0(t)D(\alpha)$, we have
\begin{IEEEeqnarray}{rCl}
\label{eq.Uret_cs}
U_\textrm{ret}(t)&=&-\frac{i}{\hbar}\theta(t)\textrm{Tr}\left( D(\alpha)\rho_0 D^\dagger(\alpha)[\hat{A}_0^\dagger(t),\hat{A}_0(0)]\right)\nonumber\\
&=&-\frac{i}{\hbar}\theta(t)\langle[\hat{A}_0(t),\hat{A}(0)]\rangle_{cs},
\end{IEEEeqnarray}
where the index $cs$ refers to the coherent states of the magnetic Hamiltonian contribution. The averages on the
normal metal are determined by using the usual Fermi-Dirac distribution.

\section{FMR-Driven spin current}
\label{sec.sp}

Once we have used the SCHA to obtain the coherent magnetization state, we can now determine the FMR-driven spin
current. Due to the dynamic field, the magnetization starts to precess, and coherent magnons fill the magnetic film 
transporting angular momentum over all directions. When a spin sink (the NM, in our case) is available, the spin current
is allowed to leak across the NM/FMI interface. We apply Eq. (\ref{eq.Uret_cs}) to Eq. (\ref{eq.Is}) to evaluate
the injected spin current, considering $\delta\mu=0$ to avoid any contribution from spin back-flow. 

The retarded Green's function is generally determined by using the Matsubara formalism \cite{mahan}. In this
case, we use the imaginary time Green's function, defined by $\hbar\mathcal{U}(\tau)=-\langle T_\tau \hat{A}(\tau)\hat{A}^\dagger(0)\rangle$, 
to make the association in the Fourier space $\mathcal{U}(i\hbar\nu_n)|_{i\hbar\nu_n\to\delta\mu+i\varepsilon}=U_\textrm{ret}(\delta\mu)$,
where $i\hbar\nu_n= n\pi/\beta$ are the bosonic (fermionic) frequencies for $n$ even (odd). 
However, the correspondence provided by the analytic continuation does not work when dealing with coherent states, and the
retarded Green's function must be solved in real time formalism. The correspondence between the Matsubara and retarded
Green formalisms fails due to the replacement of $\rho_0$ by $\rho_{cs}$. In this case, we can not use the same eigenvalues basis
for $\rho_{cs}$ and $e^{iHt}$, necessary condition to get the correspondence. Therefore the commutator present is 
$U_\textrm{ret}(t)$ is determined in real time basis and given by
\begin{widetext}
\begin{IEEEeqnarray}{rCl}
\langle[\hat{A}_0(t),\hat{A}^\dagger(0)]\rangle_{cs}&=&2\tilde{S}\sqrt{\rho_\textrm{eff}}\sum_{kk^\prime q}|\Lambda_{kk^\prime q}|^2 (f_k-f_{k^\prime})e^{\frac{i}{\hbar}(\xi_{k\uparrow}-\xi_{k^\prime\downarrow})t}\left[\sinh\theta_q\cosh\theta_q\left(\bar{\alpha}_q^2 e^{i\omega_q t}+\alpha_q^2 e^{-i\omega t}\right)+\right.\nonumber\\
&&\left.+(n_q-n_{k-k^\prime}+|\alpha_q|^2)\left(\cosh^2\theta_q e^{i\omega_q t}+\sinh^2\theta_q e^{-i\omega_q t}\right)\right],
\end{IEEEeqnarray}
\end{widetext}
where $n_q=(e^{\beta\hbar\omega_q}-1)^{-1}$ and $n_{k-k^\prime}=(e^{\beta(\xi_{k\uparrow}-\xi_{k^\prime\downarrow})}-1)^{-1}$ are 
Bose-Einstein distribution, while $f_k=(e^{\xi_{k\uparrow}}+1)^{-1}$ ($f_{k^\prime}$) is the Fermi-Dirac 
distribution for spin-up (spin-down) electrons. The second-order term 
\begin{equation}
|\Lambda_{kk^\prime q}|^2=J_{sd}^2\sum_{ij}\frac{e^{i({\bf k}-{\bf k^\prime}-{\bf q})\cdot({\bf r_j}-{\bf r_i})}}{{N_e^2N_m}}\approx\frac{J_{sd}^2N_\textrm{int}}{N_e^2N_m}
\end{equation}
is related to electrons and magnons that are created or annihilated at positions ${\bf r_i}$ and ${\bf r_j}$ on the
interface. The approximation was adopted considering that particles are created and annihilated at close positions,
and $N_\textrm{int}$ is the number of sites at the interface. 

Assuming $\delta\mu=0$, the above equation is considerably simplified, and the imaginary part of the time Fourier
transform is given by
\begin{widetext}
\begin{IEEEeqnarray}{rCl}
\textrm{Im}U_\textrm{ret}(\delta\mu=0)&=&-2\pi \tilde{S}\sqrt{\rho_\textrm{eff}}\left\{\left[|\alpha_0|^2\cosh^2\theta_0+\frac{\alpha_0^2+\bar{\alpha}_0^2}{2}\sinh\theta_0\cosh\theta_0\right]\sum_{kk^\prime}(f_k-f_{k^\prime})\delta(\epsilon_k-\epsilon_{k^\prime}+\hbar\omega_0)+\right.\nonumber\\
&&\left.+\left[|\alpha_0|^2\sinh^2\theta_0+\frac{\alpha_0^2+\bar{\alpha}_0^2}{2}\sinh\theta_0\cosh\theta_0\right]\sum_{kk^\prime}(f_k-f_{k^\prime})\delta(\epsilon_k-\epsilon_{k^\prime}-\hbar\omega_0)\right\}.
\end{IEEEeqnarray}
\end{widetext}

Considering the typical energy scale achieved in experimental arrangements, the electron momentum sum provides the simple result
\begin{equation}
\label{eq.sumkk}
\sum_{kk^\prime}(f_k-f_{k^\prime})\delta(\epsilon_k-\epsilon_{k^\prime}\pm\hbar\omega_0)\approx\pm\frac{\rho_F^2\hbar\omega_0}{4},
\end{equation}
where $\rho_F=(2m^3\epsilon_F)^{1/2}V/\pi^2\hbar^3$ is the density of states at the Fermi level 
(details can be see in Appendix \ref{appendix2}). Therefore the FMR-driven spin current density ($=I_{SP}/A$)
is given by
\begin{equation}
\label{eq.jsp_scha}
J_{SP}(T)=\left(\frac{J_{sd}\rho_F}{a N_e}\right)^2\pi\tilde{S}\hbar\omega_0|f_\alpha(T)|^2\sqrt{\rho_\textrm{eff}(T)},
\end{equation}
where we adopt a sample with interfacial area $A=L^2$. Figure \ref{fig.Isp} shows the FMR-driven spin pumping dependence on temperature. 
There is an apparent decrease with increasing temperature, which is expected provided by the reduction of magnetization coherence. 
It is important to emphasize that the temperature change is
homogeneous over the junction, and we do not take into account the Seebeck effect, which comes from temperature
gradients. In addition, above the critical temperature, SCHA predicts $\rho_E=\rho_B=0$, which will result in a vanishing FMR-driven 
spin current; however, since the system shows a paramagnetic phase, a finite spin current could be provided from EPR. Both cases, 
FM with temperature gradients and EPR-driven spin current, are fascinating problems. However, they demand a profound reformulation 
of the SCHA method, which is beyond the present work's scope.

\begin{figure}[h]
\centering \epsfig{file=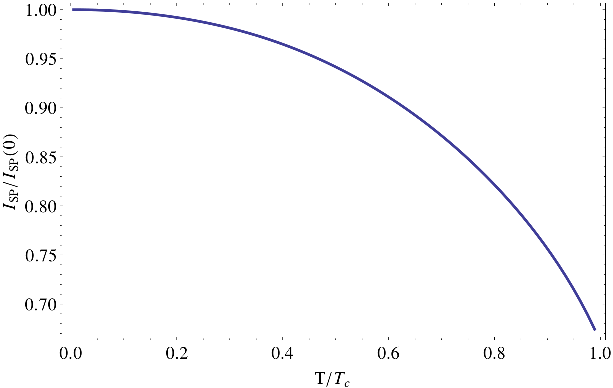,width=0.9\linewidth}
\caption{The temperature dependence of the FMR-driven spin current for $0\leq T\leq T_c$, where $T_c=1.83 J/k_B$ is
the critical temperature in which $\rho_E$ and $\rho_B$ tend to zero.}
\label{fig.Isp}
\end{figure}

To compare the SCHA outcomes with the well-known phenomenological results, we first briefly
review the LLG equation endowed with the Slonczewski term \cite{rmp77.1375}, which is given by
\begin{equation}
\dot{{\bf m}}=-\gamma{\bf m}\times{\bf B}_\textrm{eff}+\eta_0{\bf m}\times\dot{{\bf m}}+{\bm\tau},
\end{equation}
where ${\bf m}={\bf M}/M^s$ is the unity magnetization, ${\bf B}_\textrm{eff}$ is the effective magnetic field, $\eta_0$ 
is the bulk Gilbert damping, and ${\bm \tau}=(\gamma/M^s V_m)({\bf m}_\perp\times{\bf I}_{SP}\times{\bf m}_\perp)$ 
is a torque due to the angular momentum 
leaking to the NM side. The LLG equation preserves the magnetization modulus; however, since the damping
is small, we will consider an almost constant longitudinal magnetization component, while 
$\dot{{\bf m}}\approx \dot{{\bf m}}_\perp=\dot{\bf m}^y+\dot{\bf m}^z$. Another approach, which takes into account different damping
from transverse and longitudinal magnetization components, is given by the Lifshitz-Landau-Bloch-Bloembergen (LLBB) equation 
\cite{pr78.572,pr93.72,yalccin,rezende}. Provided minor corrections, the SCHA method can also be applied to the LLBB
equation as well. For a thick magnetic film, the injected spin current is related to the magnetization dynamics via
\begin{equation}
\label{eq.isp_llg}
{\bf I}_{SP}=\frac{\hbar}{4\pi}\left(G^{\uparrow\downarrow}_r{\bf m}\times\dot{\bf m}-G^{\uparrow\downarrow}_i\dot{\bf m}\right),
\end{equation}
where $G^{\uparrow\downarrow}=G^{\uparrow\downarrow}_r+iG^{\uparrow\downarrow}_i$ is 
the dimensionless spin-mixing conductance, which can be determined from the scattering-matrix theory of transport \cite{prb66.224403}. 
In general, $G^{\uparrow\downarrow}_r\gg G^{\uparrow\downarrow}_i$ and, from now on, we will consider only the real 
part of $G^{\uparrow\downarrow}$. Therefore the spin current torque results in an additional contribution to the
total Gilbert damping \cite{prl111.097602,prb89.174417,apl102.012402}, which is written as $\eta=\eta_0+\delta\eta$, where
\begin{equation}
\delta\eta=\frac{\gamma\hbar G^{\uparrow\downarrow}}{4\pi M^s V_m}
\end{equation}
is the extra damping from the FMR-driven spin current. It is more convenient to define the spin-mixing 
conductance per area $g^{\uparrow\downarrow}=G^{\uparrow\downarrow}/A$, and, for YIG films, typical spin-mixing conductance 
values are found over the interval $1.1-3.9\times 10^{18} m^{-2}$, while $\eta_0\sim 10^{-3}$, 
depending on geometric and intrinsic sample properties \cite{prl107.066604,jap111.07c307,pra1.044004}.

Using the transverse magnetization components obtained from SCHA, we get
\begin{equation}
({\bf m}\times\dot{\bf m})^x=-\frac{(\gamma B_\perp)^2\Omega (\rho_E^{1/4}+\rho_B^{1/4})^2\sqrt{\rho_\textrm{eff}}}{4\tilde{S}[(\Omega-\omega_0)^2+(\eta_0\Omega)^2]},
\end{equation}
where the minus sign is related to the clockwise direction of the precession.
Comparing the results of Eqs. ({\ref{eq.jsp_scha}) and ({\ref{eq.isp_llg}), we obtain
the spin-mixing conductance
\begin{equation}
g^{\uparrow\downarrow}=\frac{4\tilde{S}^3 J_{sd}^2 m_e^3 \epsilon_F}{\pi^2\hbar^6\rho_e^2 a^2},
\end{equation}
where $\epsilon_F\sim 10eV$ is the Fermi energy, $m_e$ is the electron mass, and $\rho_e\sim 10^{28} m^{-3}$ the
NM electron density. For $J_{sd}\sim 0.1 eV$, and $a\sim 1 nm$, the above equation provides
$g^{\uparrow\downarrow}\approx 2.6\times 10^{18}m^{-2}$, which is in remarkable agreement with experimental values. 
On the other hand, the additional Gilbert damping is given by
\begin{equation}
\delta\eta=\frac{m_e^3a J_{sd}^2 \tilde{S}^2\epsilon_F}{\pi^3\hbar^6\rho_e^2 d_m},
\end{equation}
where $d_m$ is the magnetic film thickness. For the same estimated parameters used above, and $d_m= 10^{-6}m$, 
we found $\delta\eta\approx 1.5\times 10^{-4}$, which is also within the expected.

Finally, we can also use the SCHA formalism to determine the magnetic susceptibility, which provides a valuable link
between theory and experimental measurements \cite{rezende}. Indeed, measurements from ISHE voltage in the NM side side 
directly correspond with the real and imaginary part of the magnetic susceptibility, and they are commonly used 
to get information about ferromagnetic damping. To obtain the susceptibility, we introduce the circularly polarized
magnetization
\begin{equation}
M_q^+=M_q^y+iM_q^z\approx\frac{g\mu_B}{a^3}\sqrt{2\tilde{S}}\rho_\textrm{eff}^{1/4}|\alpha_q|e^{-i(\Omega t-\phi_0)},
\end{equation}
where we replace $\rho_E$ and $\rho_B$ by $\rho_\textrm{eff}$ to simplify the result. Similarly, we define
$B_q^+(t)=\sqrt{N_m}B_\perp\delta_{q,0}e^{-i\Omega t}$, which, after a simple calculation, provides
\begin{equation}
M_q^+(t)= \frac{\gamma M^s\sqrt{\rho_\textrm{eff}}}{\Omega-\omega_0-i\eta_0\Omega}B_q^+(t).
\end{equation}
Since $B_q^+=\mu_0(H_q^++M_q^+)$, we found $M_q^+=\chi H_q^+$, with the magnetic susceptibility expressed by
\begin{equation}
\chi=\chi^\prime+i\chi^{\prime\prime}=\frac{\omega_M}{\Omega-\omega_H-i\eta_0\Omega},
\end{equation} 
where $\omega_H=\gamma\mu_0 H^x\sqrt{\rho_\textrm{eff}}$, and $\omega_M=\gamma\mu_0 M^s\sqrt{\rho_\textrm{eff}}$. Both real
and imaginary parts of $\chi$ are shown in Fig. \ref{fig.chi}. Only to provide better visualization, we chose $\eta_0=10^{-2}$, and
the vertical axis is normalized in terms of $\omega_M/\eta_0\Omega$. Apart from the renormalization factor, the magnetic
susceptibility is identical to the well-known result found in the literature. The imaginary part of the susceptibility assumes half of the
peak at the points $\omega_H=(1\pm\eta_0)\Omega$, and the difference between them is used to define the linewidth $\Delta H$
(the same linewidth also be determined by the difference between the points that define the maximum and minimum of $\chi^\prime$).
It is easy to demonstrate that $\eta_0=\sqrt{\rho_\textrm{eff}}\gamma\Delta H/\Omega$ and so, the linewidth provides
an alternative to determine the ferromagnetic damping. In addition, the imaginary part is proportional to the power
absorption (per volume) of the oscillating field by the sample. Indeed, we can show that 
$P(\Omega)=\mu_0\Omega\chi^{\prime\prime}H_\perp^2/2$, and thus the absorption radiation is maximum close to the resonant
condition $\omega_0=\Omega$.

\begin{figure}[h]
\centering \epsfig{file=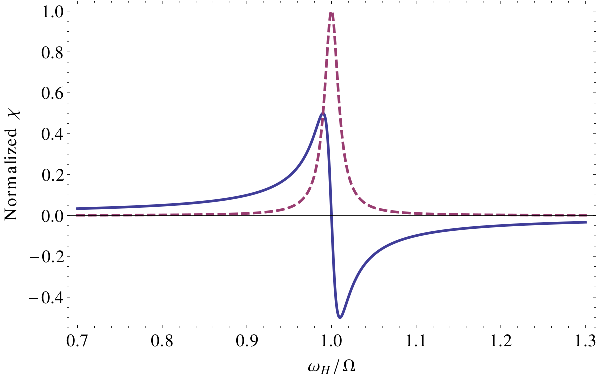,width=0.9\linewidth}
\caption{The real (solid line) and imaginary (dashed line) parts of the magnetic susceptibility 
(normalized in terms of $\omega_M/\eta_0\Omega$) obtained from the SCHA formalism.}
\label{fig.chi}
\end{figure}

\section{Summary and conclusions}
\label{sec.conclusion}

In this work, we applied the SCHA formalism and coherent states to investigate the FMR-driven spin current in an NM/FMI junction. 
Over the years, similar problems have been analyzed through bosonic representations or phenomenological
approaches. Provided the coherent nature of ferromagnetic resonance, it is appropriated 
to apply the coherent state formalism. In addition, in the SCHA formalism, the entire development is performed through
the $\varphi$ and $S^z$ operators that satisfy $[\varphi_i,S_j^z]=\delta_{ij}$. Thus the SCHA is
an advantageous method for studying problems involving FMR.

Here, we considered the application of a resonant driving field in an NM/FMI junction to provide the injection of
spin current into the normal metal. First, the FMR-driven spin pumping was determined by using an sd coupling at the
interface. Then, a precise and detailed development was performed to obtain, beyond the spin pumping current, the
spin-mixing conductance, the extra magnetic damping from the spin pumping, and the susceptibility. The SCHA 
results showed considerable agreement with experimental data when considering typical experimental values of the 
involved parameters. 

In summary, we have demonstrated the efficiency of the SCHA method, combined with coherent states, to treat 
magnetic problems in spintronics. Therefore a series of open problems could also be investigated by using the presented 
formalism that would result in a breakthrough for much spintronic research. \\

\appendix

\section{Renormalization parameteres}
\label{appendix1}
To determine the renormalization parameter, we compare the value of $\langle\dot{S}_q^z \dot{S}_{-q}^z\rangle_0$
obtained from the quadratic Hamiltonian \ref{eq.H0m} with the result obtained from $H^m$ without the series
expansion. Starting with the former, and considering the semiclassical analysis, 	we obtain:
\begin{equation}
\langle\hbar^2\dot{S}_q^z\dot{S}_{-q}^z\rangle_0=(h_q^\varphi\tilde{S}^2)^2\langle\varphi_q\varphi_{-q}\rangle_0=\frac{h_q^\varphi\tilde{S}^2}{\beta},
\end{equation}
with $h_q^\varphi=g\mu_B B^x\rho_B+zJ\tilde{S}^2(1-\gamma_q)\rho_E$. To find out the second term, given by the Fourier transform
\begin{equation}
\langle \dot{S}_q^z\dot{S}_{-q}^z\rangle=\frac{1}{N}\sum_{ij}\langle\dot{S}_i^z\dot{S}_j^z\rangle e^{i{\bf q}\cdot({\bf r}_j-{\bf r}_i)},
\end{equation}
we use the following useful relation, obtained after an integration by parts,
\begin{equation}
\langle\hbar^2\dot{S}_i^z\dot{S}_j^z\rangle=\frac{1}{Z}\int\mathcal{D}\varphi\mathcal{D}S^z\frac{1}{\beta}\frac{\partial^2H^m}{\partial\varphi_i\partial\varphi_j}e^{-\beta H^m},
\end{equation}
where $Z$ is the partition function, and the integration measure $\mathcal{D}\varphi\mathcal{D}S^z$ stands for the field
integration over each site on the lattice. In addition, we extend the integration limit to $-\infty<\varphi,S^z<\infty$
and thus we deal with Gaussian integrals. The derivative of the semiclassical Hamiltonian provides
\begin{IEEEeqnarray}{l}
\frac{\partial^2H^m}{\partial\varphi_i\partial\varphi_j}=g\mu_B B^x\sqrt{\tilde{S}^2-(S_i^z)^2}\cos\varphi_i\delta_{ij}+\nonumber\\
+J\sum_l\sqrt{\tilde{S}^2-(S_i^z)^2}\sqrt{\tilde{S}^2-(S_l^z)^2}\cos(\varphi_l-\varphi_j)(\delta_{ij}-\delta_{il}).\nonumber
\end{IEEEeqnarray}
The Fourier transform is then written as
\begin{IEEEeqnarray}{l}
\langle\hbar^2\dot{S}_q^z\dot{S}_{-q}^z\rangle=\frac{g\mu_B B^x}{\beta}\langle\sqrt{\tilde{S}^2-(S^z)^2}\cos\varphi\rangle+\nonumber\\
+\frac{zJ(1-\gamma_q)}{\beta}\langle[\tilde{S}^2-(S^z)^2]\cos\Delta\varphi\rangle,
\end{IEEEeqnarray}
where we consider that the averages are site independent. Comparing with the previous result and 
using the decoupled quadratic Hamiltonian $H_0^m$ to evaluate the average, we obtain the self-consistent equations used in the text.

\section{Electron momentum sum}
\label{appendix2}
In order to evaluate the momentum sum of up- and down-spin, we use the conservation energy condition to write
\begin{equation}
\sum_k \delta(\epsilon_k-\epsilon_{k^\prime}\pm\hbar\omega_0)\approx\frac{mV}{2\pi^2\hbar^2}k^\prime\left(1\pm\frac{\hbar\omega_0}{2\epsilon_{k^\prime}}\right),
\end{equation}
and thus, in the continuum limit, the left-hand side (l.h.s) of Eq. (\ref{eq.sumkk}) is given by
\begin{IEEEeqnarray}{l}
\label{eq.app1}
\frac{m^2 V^2}{4\pi^3\beta\hbar^4}\left\{\frac{2m}{\beta\hbar^2}[F_1(\beta\mu\pm\beta\hbar\omega_0)-F_1(\beta\mu)]\pm\right.\nonumber\\
\left.\pm\frac{m\omega_0}{\hbar}[F_0(\beta\mu\pm\beta\hbar\omega_0)-F_0(\beta\mu)]\right\},
\end{IEEEeqnarray}
where the integral was written as
\begin{equation}
\int\frac{d^3 k^\prime}{(2\pi)^3}\frac{f(\xi\pm\hbar\omega_0)}{k^{3-2l}}=\frac{\Gamma(l)}{4\pi}\left(\frac{2m}{\beta\hbar^2}\right)^l F_{l-1}(\beta\mu\mp\beta\hbar\omega_0)
\end{equation}
with
\begin{equation}
F_l(x)=\frac{1}{\Gamma(l+1)}\int_0^\infty dz z^l(e^{z-x}+1)^{-1}
\end{equation}
being the complete Fermi-Dirac integral. Here, we have considered the perfect spin sink limit,
{\it i.e.}, $\mu_\uparrow=\mu_\downarrow=\mu$. For $l=0$, we have the exact result $F_0(x)=\ln(1+e^x)$, while the derivatives 
are given by $dF_l/dx=F_{l-1}(x)$. In usual FMR experiments, we deal with the energies $\mu\approx\epsilon_F\sim 10 eV$, $k_B T\sim 10^{-2} eV$,
and $\hbar\omega_0\sim 10^{-6} eV$, which justify a Taylor expansion of the Fermi-Dirac integral $F_1$ around $\beta\mu$ ($F_0$ is
assumed constant). Therefore Eq. (\ref{eq.app1}) provides
\begin{equation}
\sum_{kk^\prime}(f_k-f_{k^\prime})\delta(\epsilon_k-\epsilon_{k^\prime}\pm\hbar\omega_0)=\frac{m^3 V^2}{2\pi^4\hbar^6}\frac{F_0(\beta\mu)}{\beta},
\end{equation}
which directly results in Eq. (\ref{eq.sumkk}).

\bibliography{manuscript}

\begin{thebibliography}{77}%
\makeatletter
\providecommand \@ifxundefined [1]{%
 \@ifx{#1\undefined}
}%
\providecommand \@ifnum [1]{%
 \ifnum #1\expandafter \@firstoftwo
 \else \expandafter \@secondoftwo
 \fi
}%
\providecommand \@ifx [1]{%
 \ifx #1\expandafter \@firstoftwo
 \else \expandafter \@secondoftwo
 \fi
}%
\providecommand \natexlab [1]{#1}%
\providecommand \enquote  [1]{``#1''}%
\providecommand \bibnamefont  [1]{#1}%
\providecommand \bibfnamefont [1]{#1}%
\providecommand \citenamefont [1]{#1}%
\providecommand \href@noop [0]{\@secondoftwo}%
\providecommand \href [0]{\begingroup \@sanitize@url \@href}%
\providecommand \@href[1]{\@@startlink{#1}\@@href}%
\providecommand \@@href[1]{\endgroup#1\@@endlink}%
\providecommand \@sanitize@url [0]{\catcode `\\12\catcode `\$12\catcode
  `\&12\catcode `\#12\catcode `\^12\catcode `\_12\catcode `\%12\relax}%
\providecommand \@@startlink[1]{}%
\providecommand \@@endlink[0]{}%
\providecommand \url  [0]{\begingroup\@sanitize@url \@url }%
\providecommand \@url [1]{\endgroup\@href {#1}{\urlprefix }}%
\providecommand \urlprefix  [0]{URL }%
\providecommand \Eprint [0]{\href }%
\providecommand \doibase [0]{http://dx.doi.org/}%
\providecommand \selectlanguage [0]{\@gobble}%
\providecommand \bibinfo  [0]{\@secondoftwo}%
\providecommand \bibfield  [0]{\@secondoftwo}%
\providecommand \translation [1]{[#1]}%
\providecommand \BibitemOpen [0]{}%
\providecommand \bibitemStop [0]{}%
\providecommand \bibitemNoStop [0]{.\EOS\space}%
\providecommand \EOS [0]{\spacefactor3000\relax}%
\providecommand \BibitemShut  [1]{\csname bibitem#1\endcsname}%
\let\auto@bib@innerbib\@empty
%</preamble>
\bibitem [{\citenamefont {Wolf}\ \emph {et~al.}(2001)\citenamefont {Wolf},
  \citenamefont {Awschalom}, \citenamefont {Buhrman}, \citenamefont {Daughton},
  \citenamefont {Von~Molnar}, \citenamefont {Roukes}, \citenamefont
  {Chtchelkanova},\ and\ \citenamefont {Treger}}]{science294.1488}%
  \BibitemOpen
  \bibfield  {author} {\bibinfo {author} {\bibfnamefont {S.}~\bibnamefont
  {Wolf}}, \bibinfo {author} {\bibfnamefont {D.}~\bibnamefont {Awschalom}},
  \bibinfo {author} {\bibfnamefont {R.}~\bibnamefont {Buhrman}}, \bibinfo
  {author} {\bibfnamefont {J.}~\bibnamefont {Daughton}}, \bibinfo {author}
  {\bibfnamefont {S.}~\bibnamefont {Von~Molnar}}, \bibinfo {author}
  {\bibfnamefont {M.}~\bibnamefont {Roukes}}, \bibinfo {author} {\bibfnamefont
  {A.~Y.}\ \bibnamefont {Chtchelkanova}}, \ and\ \bibinfo {author}
  {\bibfnamefont {D.}~\bibnamefont {Treger}},\ }\href@noop {} {\bibfield
  {journal} {\bibinfo  {journal} {Science}\ }\textbf {\bibinfo {volume}
  {294}},\ \bibinfo {pages} {1488} (\bibinfo {year} {2001})}\BibitemShut
  {NoStop}%
\bibitem [{\citenamefont {{\v{Z}}uti{\'c}}\ \emph {et~al.}(2004)\citenamefont
  {{\v{Z}}uti{\'c}}, \citenamefont {Fabian},\ and\ \citenamefont
  {Sarma}}]{rmp76.323}%
  \BibitemOpen
  \bibfield  {author} {\bibinfo {author} {\bibfnamefont {I.}~\bibnamefont
  {{\v{Z}}uti{\'c}}}, \bibinfo {author} {\bibfnamefont {J.}~\bibnamefont
  {Fabian}}, \ and\ \bibinfo {author} {\bibfnamefont {S.~D.}\ \bibnamefont
  {Sarma}},\ }\href@noop {} {\bibfield  {journal} {\bibinfo  {journal} {Reviews
  of Modern Physics}\ }\textbf {\bibinfo {volume} {76}},\ \bibinfo {pages}
  {323} (\bibinfo {year} {2004})}\BibitemShut {NoStop}%
\bibitem [{\citenamefont {Hirsch}(1999)}]{prl83.1834}%
  \BibitemOpen
  \bibfield  {author} {\bibinfo {author} {\bibfnamefont {J.~E.}\ \bibnamefont
  {Hirsch}},\ }\href@noop {} {\bibfield  {journal} {\bibinfo  {journal}
  {Physical Review Letters}\ }\textbf {\bibinfo {volume} {83}},\ \bibinfo
  {pages} {1834} (\bibinfo {year} {1999})}\BibitemShut {NoStop}%
\bibitem [{\citenamefont {Rezende}(2020)}]{rezende}%
  \BibitemOpen
  \bibfield  {author} {\bibinfo {author} {\bibfnamefont {S.~M.}\ \bibnamefont
  {Rezende}},\ }\href@noop {} {\emph {\bibinfo {title} {Fundamentals of
  Magnonics}}},\ Vol.\ \bibinfo {volume} {969}\ (\bibinfo  {publisher}
  {Springer},\ \bibinfo {address} {Switzerland},\ \bibinfo {year}
  {2020})\BibitemShut {NoStop}%
\bibitem [{\citenamefont {Hirobe}\ \emph {et~al.}(2017)\citenamefont {Hirobe},
  \citenamefont {Sato}, \citenamefont {Kawamata}, \citenamefont {Shiomi},
  \citenamefont {Uchida}, \citenamefont {Iguchi}, \citenamefont {Koike},
  \citenamefont {Maekawa},\ and\ \citenamefont {Saitoh}}]{naturephys13.30}%
  \BibitemOpen
  \bibfield  {author} {\bibinfo {author} {\bibfnamefont {D.}~\bibnamefont
  {Hirobe}}, \bibinfo {author} {\bibfnamefont {M.}~\bibnamefont {Sato}},
  \bibinfo {author} {\bibfnamefont {T.}~\bibnamefont {Kawamata}}, \bibinfo
  {author} {\bibfnamefont {Y.}~\bibnamefont {Shiomi}}, \bibinfo {author}
  {\bibfnamefont {K.-i.}\ \bibnamefont {Uchida}}, \bibinfo {author}
  {\bibfnamefont {R.}~\bibnamefont {Iguchi}}, \bibinfo {author} {\bibfnamefont
  {Y.}~\bibnamefont {Koike}}, \bibinfo {author} {\bibfnamefont
  {S.}~\bibnamefont {Maekawa}}, \ and\ \bibinfo {author} {\bibfnamefont
  {E.}~\bibnamefont {Saitoh}},\ }\href@noop {} {\bibfield  {journal} {\bibinfo
  {journal} {Nature Physics}\ }\textbf {\bibinfo {volume} {13}},\ \bibinfo
  {pages} {30} (\bibinfo {year} {2017})}\BibitemShut {NoStop}%
\bibitem [{\citenamefont {Lange}\ \emph {et~al.}(2018)\citenamefont {Lange},
  \citenamefont {Ejima}, \citenamefont {Shirakawa}, \citenamefont {Yunoki},\
  and\ \citenamefont {Fehske}}]{prb97.245124}%
  \BibitemOpen
  \bibfield  {author} {\bibinfo {author} {\bibfnamefont {F.}~\bibnamefont
  {Lange}}, \bibinfo {author} {\bibfnamefont {S.}~\bibnamefont {Ejima}},
  \bibinfo {author} {\bibfnamefont {T.}~\bibnamefont {Shirakawa}}, \bibinfo
  {author} {\bibfnamefont {S.}~\bibnamefont {Yunoki}}, \ and\ \bibinfo {author}
  {\bibfnamefont {H.}~\bibnamefont {Fehske}},\ }\href@noop {} {\bibfield
  {journal} {\bibinfo  {journal} {Physical Review B}\ }\textbf {\bibinfo
  {volume} {97}},\ \bibinfo {pages} {245124} (\bibinfo {year}
  {2018})}\BibitemShut {NoStop}%
\bibitem [{\citenamefont {Hirobe}\ \emph {et~al.}(2018)\citenamefont {Hirobe},
  \citenamefont {Kawamata}, \citenamefont {Oyanagi}, \citenamefont {Koike},\
  and\ \citenamefont {Saitoh}}]{jap123.123903}%
  \BibitemOpen
  \bibfield  {author} {\bibinfo {author} {\bibfnamefont {D.}~\bibnamefont
  {Hirobe}}, \bibinfo {author} {\bibfnamefont {T.}~\bibnamefont {Kawamata}},
  \bibinfo {author} {\bibfnamefont {K.}~\bibnamefont {Oyanagi}}, \bibinfo
  {author} {\bibfnamefont {Y.}~\bibnamefont {Koike}}, \ and\ \bibinfo {author}
  {\bibfnamefont {E.}~\bibnamefont {Saitoh}},\ }\href@noop {} {\bibfield
  {journal} {\bibinfo  {journal} {Journal of Applied Physics}\ }\textbf
  {\bibinfo {volume} {123}},\ \bibinfo {pages} {123903} (\bibinfo {year}
  {2018})}\BibitemShut {NoStop}%
\bibitem [{\citenamefont {Slonczewski}(1996)}]{jmmm159.l1}%
  \BibitemOpen
  \bibfield  {author} {\bibinfo {author} {\bibfnamefont {J.~C.}\ \bibnamefont
  {Slonczewski}},\ }\href@noop {} {\bibfield  {journal} {\bibinfo  {journal}
  {Journal of Magnetism and Magnetic Materials}\ }\textbf {\bibinfo {volume}
  {159}},\ \bibinfo {pages} {L1} (\bibinfo {year} {1996})}\BibitemShut
  {NoStop}%
\bibitem [{\citenamefont {Berger}(1996)}]{prb54.9353}%
  \BibitemOpen
  \bibfield  {author} {\bibinfo {author} {\bibfnamefont {L.}~\bibnamefont
  {Berger}},\ }\href@noop {} {\bibfield  {journal} {\bibinfo  {journal}
  {Physical Review B}\ }\textbf {\bibinfo {volume} {54}},\ \bibinfo {pages}
  {9353} (\bibinfo {year} {1996})}\BibitemShut {NoStop}%
\bibitem [{\citenamefont {Tserkovnyak}\ \emph
  {et~al.}(2002{\natexlab{a}})\citenamefont {Tserkovnyak}, \citenamefont
  {Brataas},\ and\ \citenamefont {Bauer}}]{prl88.117601}%
  \BibitemOpen
  \bibfield  {author} {\bibinfo {author} {\bibfnamefont {Y.}~\bibnamefont
  {Tserkovnyak}}, \bibinfo {author} {\bibfnamefont {A.}~\bibnamefont
  {Brataas}}, \ and\ \bibinfo {author} {\bibfnamefont {G.~E.~W.}\ \bibnamefont
  {Bauer}},\ }\href@noop {} {\bibfield  {journal} {\bibinfo  {journal}
  {Physical Review Letters}\ }\textbf {\bibinfo {volume} {88}},\ \bibinfo
  {pages} {117601} (\bibinfo {year} {2002}{\natexlab{a}})}\BibitemShut
  {NoStop}%
\bibitem [{\citenamefont {Shiomi}\ and\ \citenamefont
  {Saitoh}(2014)}]{prl113.266602}%
  \BibitemOpen
  \bibfield  {author} {\bibinfo {author} {\bibfnamefont {Y.}~\bibnamefont
  {Shiomi}}\ and\ \bibinfo {author} {\bibfnamefont {E.}~\bibnamefont
  {Saitoh}},\ }\href@noop {} {\bibfield  {journal} {\bibinfo  {journal}
  {Physical Review Letters}\ }\textbf {\bibinfo {volume} {113}},\ \bibinfo
  {pages} {266602} (\bibinfo {year} {2014})}\BibitemShut {NoStop}%
\bibitem [{\citenamefont {Azevedo}\ \emph {et~al.}(2005)\citenamefont
  {Azevedo}, \citenamefont {Vilela~Leao}, \citenamefont {Rodriguez-Suarez},
  \citenamefont {Oliveira},\ and\ \citenamefont {Rezende}}]{jap97.10c715}%
  \BibitemOpen
  \bibfield  {author} {\bibinfo {author} {\bibfnamefont {A.}~\bibnamefont
  {Azevedo}}, \bibinfo {author} {\bibfnamefont {L.}~\bibnamefont
  {Vilela~Leao}}, \bibinfo {author} {\bibfnamefont {R.}~\bibnamefont
  {Rodriguez-Suarez}}, \bibinfo {author} {\bibfnamefont {A.}~\bibnamefont
  {Oliveira}}, \ and\ \bibinfo {author} {\bibfnamefont {S.}~\bibnamefont
  {Rezende}},\ }\href@noop {} {\bibfield  {journal} {\bibinfo  {journal}
  {Journal of Applied Physics}\ }\textbf {\bibinfo {volume} {97}},\ \bibinfo
  {pages} {10C715} (\bibinfo {year} {2005})}\BibitemShut {NoStop}%
\bibitem [{\citenamefont {Saitoh}\ \emph {et~al.}(2006)\citenamefont {Saitoh},
  \citenamefont {Ueda}, \citenamefont {Miyajima},\ and\ \citenamefont
  {Tatara}}]{apl88.182509}%
  \BibitemOpen
  \bibfield  {author} {\bibinfo {author} {\bibfnamefont {E.}~\bibnamefont
  {Saitoh}}, \bibinfo {author} {\bibfnamefont {M.}~\bibnamefont {Ueda}},
  \bibinfo {author} {\bibfnamefont {H.}~\bibnamefont {Miyajima}}, \ and\
  \bibinfo {author} {\bibfnamefont {G.}~\bibnamefont {Tatara}},\ }\href@noop {}
  {\bibfield  {journal} {\bibinfo  {journal} {Applied Physics Letters}\
  }\textbf {\bibinfo {volume} {88}},\ \bibinfo {pages} {182509} (\bibinfo
  {year} {2006})}\BibitemShut {NoStop}%
\bibitem [{\citenamefont {Brataas}\ \emph {et~al.}(2012)\citenamefont
  {Brataas}, \citenamefont {Tserkovnyak}, \citenamefont {Bauer},\ and\
  \citenamefont {Kelly}}]{spincurrent12.87}%
  \BibitemOpen
  \bibfield  {author} {\bibinfo {author} {\bibfnamefont {A.}~\bibnamefont
  {Brataas}}, \bibinfo {author} {\bibfnamefont {Y.}~\bibnamefont
  {Tserkovnyak}}, \bibinfo {author} {\bibfnamefont {G.}~\bibnamefont {Bauer}},
  \ and\ \bibinfo {author} {\bibfnamefont {P.~J.}\ \bibnamefont {Kelly}},\
  }\href@noop {} {\bibfield  {journal} {\bibinfo  {journal} {Spin current}\
  }\textbf {\bibinfo {volume} {17}},\ \bibinfo {pages} {87} (\bibinfo {year}
  {2012})}\BibitemShut {NoStop}%
\bibitem [{\citenamefont {Gilbert}(2004)}]{ieee6.3443}%
  \BibitemOpen
  \bibfield  {author} {\bibinfo {author} {\bibfnamefont {T.~L.}\ \bibnamefont
  {Gilbert}},\ }\href@noop {} {\bibfield  {journal} {\bibinfo  {journal} {IEEE
  Transactions on Magnetics}\ }\textbf {\bibinfo {volume} {40}},\ \bibinfo
  {pages} {3443} (\bibinfo {year} {2004})}\BibitemShut {NoStop}%
\bibitem [{\citenamefont {Urban}\ \emph {et~al.}(2001)\citenamefont {Urban},
  \citenamefont {Woltersdorf},\ and\ \citenamefont {Heinrich}}]{prl87.217204}%
  \BibitemOpen
  \bibfield  {author} {\bibinfo {author} {\bibfnamefont {R.}~\bibnamefont
  {Urban}}, \bibinfo {author} {\bibfnamefont {G.}~\bibnamefont {Woltersdorf}},
  \ and\ \bibinfo {author} {\bibfnamefont {B.}~\bibnamefont {Heinrich}},\
  }\href@noop {} {\bibfield  {journal} {\bibinfo  {journal} {Physical review
  letters}\ }\textbf {\bibinfo {volume} {87}},\ \bibinfo {pages} {217204}
  (\bibinfo {year} {2001})}\BibitemShut {NoStop}%
\bibitem [{\citenamefont {Tserkovnyak}\ \emph
  {et~al.}(2002{\natexlab{b}})\citenamefont {Tserkovnyak}, \citenamefont
  {Brataas},\ and\ \citenamefont {Bauer}}]{prb66.224403}%
  \BibitemOpen
  \bibfield  {author} {\bibinfo {author} {\bibfnamefont {Y.}~\bibnamefont
  {Tserkovnyak}}, \bibinfo {author} {\bibfnamefont {A.}~\bibnamefont
  {Brataas}}, \ and\ \bibinfo {author} {\bibfnamefont {G.~E.~W.}\ \bibnamefont
  {Bauer}},\ }\href@noop {} {\bibfield  {journal} {\bibinfo  {journal}
  {Physical Review B}\ }\textbf {\bibinfo {volume} {66}},\ \bibinfo {pages}
  {224403} (\bibinfo {year} {2002}{\natexlab{b}})}\BibitemShut {NoStop}%
\bibitem [{\citenamefont {Brataas}\ \emph {et~al.}(2008)\citenamefont
  {Brataas}, \citenamefont {Tserkovnyak},\ and\ \citenamefont
  {Bauer}}]{prl101.037207}%
  \BibitemOpen
  \bibfield  {author} {\bibinfo {author} {\bibfnamefont {A.}~\bibnamefont
  {Brataas}}, \bibinfo {author} {\bibfnamefont {Y.}~\bibnamefont
  {Tserkovnyak}}, \ and\ \bibinfo {author} {\bibfnamefont {G.~E.~W.}\
  \bibnamefont {Bauer}},\ }\href@noop {} {\bibfield  {journal} {\bibinfo
  {journal} {Physical review letters}\ }\textbf {\bibinfo {volume} {101}},\
  \bibinfo {pages} {037207} (\bibinfo {year} {2008})}\BibitemShut {NoStop}%
\bibitem [{\citenamefont {Kapelrud}\ and\ \citenamefont
  {Brataas}(2013)}]{prl111.097602}%
  \BibitemOpen
  \bibfield  {author} {\bibinfo {author} {\bibfnamefont {A.}~\bibnamefont
  {Kapelrud}}\ and\ \bibinfo {author} {\bibfnamefont {A.}~\bibnamefont
  {Brataas}},\ }\href@noop {} {\bibfield  {journal} {\bibinfo  {journal}
  {Physical review letters}\ }\textbf {\bibinfo {volume} {111}},\ \bibinfo
  {pages} {097602} (\bibinfo {year} {2013})}\BibitemShut {NoStop}%
\bibitem [{\citenamefont {Takahashi}\ \emph {et~al.}(2010)\citenamefont
  {Takahashi}, \citenamefont {Saitoh},\ and\ \citenamefont
  {Maekawa}}]{jpcs200.062030}%
  \BibitemOpen
  \bibfield  {author} {\bibinfo {author} {\bibfnamefont {S.}~\bibnamefont
  {Takahashi}}, \bibinfo {author} {\bibfnamefont {E.}~\bibnamefont {Saitoh}}, \
  and\ \bibinfo {author} {\bibfnamefont {S.}~\bibnamefont {Maekawa}},\
  }\bibfield  {booktitle} {\emph {\bibinfo {booktitle} {Journal of Physics:
  Conference Series}},\ }\href@noop {} {\ \textbf {\bibinfo {volume} {200}},\
  \bibinfo {pages} {062030} (\bibinfo {year} {2010})}\BibitemShut {NoStop}%
\bibitem [{\citenamefont {Ohnuma}\ \emph {et~al.}(2014)\citenamefont {Ohnuma},
  \citenamefont {Adachi}, \citenamefont {Saitoh},\ and\ \citenamefont
  {Maekawa}}]{prb89.174417}%
  \BibitemOpen
  \bibfield  {author} {\bibinfo {author} {\bibfnamefont {Y.}~\bibnamefont
  {Ohnuma}}, \bibinfo {author} {\bibfnamefont {H.}~\bibnamefont {Adachi}},
  \bibinfo {author} {\bibfnamefont {E.}~\bibnamefont {Saitoh}}, \ and\ \bibinfo
  {author} {\bibfnamefont {S.}~\bibnamefont {Maekawa}},\ }\href@noop {}
  {\bibfield  {journal} {\bibinfo  {journal} {Physical Review B}\ }\textbf
  {\bibinfo {volume} {89}},\ \bibinfo {pages} {174417} (\bibinfo {year}
  {2014})}\BibitemShut {NoStop}%
\bibitem [{\citenamefont {Okamoto}(2016)}]{prb93.064421}%
  \BibitemOpen
  \bibfield  {author} {\bibinfo {author} {\bibfnamefont {S.}~\bibnamefont
  {Okamoto}},\ }\href@noop {} {\bibfield  {journal} {\bibinfo  {journal}
  {Physical Review B}\ }\textbf {\bibinfo {volume} {93}},\ \bibinfo {pages}
  {064421} (\bibinfo {year} {2016})}\BibitemShut {NoStop}%
\bibitem [{\citenamefont {Vargas}\ and\ \citenamefont
  {Moura}(2020)}]{prb102.024412}%
  \BibitemOpen
  \bibfield  {author} {\bibinfo {author} {\bibfnamefont {V.~S. U.~A.}\
  \bibnamefont {Vargas}}\ and\ \bibinfo {author} {\bibfnamefont {A.~R.}\
  \bibnamefont {Moura}},\ }\href@noop {} {\bibfield  {journal} {\bibinfo
  {journal} {Physical Review B}\ }\textbf {\bibinfo {volume} {102}},\ \bibinfo
  {pages} {024412} (\bibinfo {year} {2020})}\BibitemShut {NoStop}%
\bibitem [{\citenamefont {Holstein}\ and\ \citenamefont
  {Primakoff}(1940)}]{pr58.1098}%
  \BibitemOpen
  \bibfield  {author} {\bibinfo {author} {\bibfnamefont {T.}~\bibnamefont
  {Holstein}}\ and\ \bibinfo {author} {\bibfnamefont {H.}~\bibnamefont
  {Primakoff}},\ }\href@noop {} {\bibfield  {journal} {\bibinfo  {journal}
  {Physical Review}\ }\textbf {\bibinfo {volume} {58}},\ \bibinfo {pages}
  {1098} (\bibinfo {year} {1940})}\BibitemShut {NoStop}%
\bibitem [{\citenamefont {Auerbach}(2012)}]{auerbach}%
  \BibitemOpen
  \bibfield  {author} {\bibinfo {author} {\bibfnamefont {A.}~\bibnamefont
  {Auerbach}},\ }\href@noop {} {\emph {\bibinfo {title} {Interacting Electrons
  and Quantum Magnetism}}}\ (\bibinfo  {publisher} {Springer Science \&
  Business Media},\ \bibinfo {address} {United States of America},\ \bibinfo
  {year} {2012})\BibitemShut {NoStop}%
\bibitem [{\citenamefont {Arovas}\ and\ \citenamefont
  {Auerbach}(1988)}]{prb38.316}%
  \BibitemOpen
  \bibfield  {author} {\bibinfo {author} {\bibfnamefont {D.~P.}\ \bibnamefont
  {Arovas}}\ and\ \bibinfo {author} {\bibfnamefont {A.}~\bibnamefont
  {Auerbach}},\ }\href@noop {} {\bibfield  {journal} {\bibinfo  {journal}
  {Physical Review B}\ }\textbf {\bibinfo {volume} {38}},\ \bibinfo {pages}
  {316} (\bibinfo {year} {1988})}\BibitemShut {NoStop}%
\bibitem [{\citenamefont {Sarker}\ \emph {et~al.}(1989)\citenamefont {Sarker},
  \citenamefont {Jayaprakash}, \citenamefont {Krishnamurthy},\ and\
  \citenamefont {Ma}}]{prb40.5028}%
  \BibitemOpen
  \bibfield  {author} {\bibinfo {author} {\bibfnamefont {S.}~\bibnamefont
  {Sarker}}, \bibinfo {author} {\bibfnamefont {C.}~\bibnamefont {Jayaprakash}},
  \bibinfo {author} {\bibfnamefont {H.~R.}\ \bibnamefont {Krishnamurthy}}, \
  and\ \bibinfo {author} {\bibfnamefont {M.}~\bibnamefont {Ma}},\ }\href@noop
  {} {\bibfield  {journal} {\bibinfo  {journal} {Physical Review B}\ }\textbf
  {\bibinfo {volume} {40}},\ \bibinfo {pages} {5028} (\bibinfo {year}
  {1989})}\BibitemShut {NoStop}%
\bibitem [{\citenamefont {Trumper}\ \emph {et~al.}(1997)\citenamefont
  {Trumper}, \citenamefont {Manuel}, \citenamefont {Gazza},\ and\ \citenamefont
  {Ceccatto}}]{prl78.2216}%
  \BibitemOpen
  \bibfield  {author} {\bibinfo {author} {\bibfnamefont {A.~E.}\ \bibnamefont
  {Trumper}}, \bibinfo {author} {\bibfnamefont {L.~O.}\ \bibnamefont {Manuel}},
  \bibinfo {author} {\bibfnamefont {C.~J.}\ \bibnamefont {Gazza}}, \ and\
  \bibinfo {author} {\bibfnamefont {H.~A.}\ \bibnamefont {Ceccatto}},\
  }\href@noop {} {\bibfield  {journal} {\bibinfo  {journal} {Physical review
  letters}\ }\textbf {\bibinfo {volume} {78}},\ \bibinfo {pages} {2216}
  (\bibinfo {year} {1997})}\BibitemShut {NoStop}%
\bibitem [{\citenamefont {Gonzalez}\ \emph {et~al.}(2017)\citenamefont
  {Gonzalez}, \citenamefont {Ghioldi}, \citenamefont {Gazza}, \citenamefont
  {Manuel},\ and\ \citenamefont {Trumper}}]{prb96.174423}%
  \BibitemOpen
  \bibfield  {author} {\bibinfo {author} {\bibfnamefont {M.~G.}\ \bibnamefont
  {Gonzalez}}, \bibinfo {author} {\bibfnamefont {E.~A.}\ \bibnamefont
  {Ghioldi}}, \bibinfo {author} {\bibfnamefont {C.~J.}\ \bibnamefont {Gazza}},
  \bibinfo {author} {\bibfnamefont {L.~O.}\ \bibnamefont {Manuel}}, \ and\
  \bibinfo {author} {\bibfnamefont {A.~E.}\ \bibnamefont {Trumper}},\
  }\href@noop {} {\bibfield  {journal} {\bibinfo  {journal} {Physical Review
  B}\ }\textbf {\bibinfo {volume} {96}},\ \bibinfo {pages} {174423} (\bibinfo
  {year} {2017})}\BibitemShut {NoStop}%
\bibitem [{\citenamefont {Ghioldi}\ \emph {et~al.}(2018)\citenamefont
  {Ghioldi}, \citenamefont {Gonzalez}, \citenamefont {Zhang}, \citenamefont
  {Kamiya}, \citenamefont {Manuel}, \citenamefont {Trumper},\ and\
  \citenamefont {Batista}}]{prb98.184403}%
  \BibitemOpen
  \bibfield  {author} {\bibinfo {author} {\bibfnamefont {E.~A.}\ \bibnamefont
  {Ghioldi}}, \bibinfo {author} {\bibfnamefont {M.~G.}\ \bibnamefont
  {Gonzalez}}, \bibinfo {author} {\bibfnamefont {S.-S.}\ \bibnamefont {Zhang}},
  \bibinfo {author} {\bibfnamefont {Y.}~\bibnamefont {Kamiya}}, \bibinfo
  {author} {\bibfnamefont {L.~O.}\ \bibnamefont {Manuel}}, \bibinfo {author}
  {\bibfnamefont {A.~E.}\ \bibnamefont {Trumper}}, \ and\ \bibinfo {author}
  {\bibfnamefont {C.~D.}\ \bibnamefont {Batista}},\ }\href@noop {} {\bibfield
  {journal} {\bibinfo  {journal} {Physical Review B}\ }\textbf {\bibinfo
  {volume} {98}},\ \bibinfo {pages} {184403} (\bibinfo {year}
  {2018})}\BibitemShut {NoStop}%
\bibitem [{\citenamefont {Zhang}\ \emph {et~al.}(2019)\citenamefont {Zhang},
  \citenamefont {Ghioldi}, \citenamefont {Kamiya}, \citenamefont {Manuel},
  \citenamefont {Trumper},\ and\ \citenamefont {Batista}}]{prb100.104431}%
  \BibitemOpen
  \bibfield  {author} {\bibinfo {author} {\bibfnamefont {S.-S.}\ \bibnamefont
  {Zhang}}, \bibinfo {author} {\bibfnamefont {E.~A.}\ \bibnamefont {Ghioldi}},
  \bibinfo {author} {\bibfnamefont {Y.}~\bibnamefont {Kamiya}}, \bibinfo
  {author} {\bibfnamefont {L.~O.}\ \bibnamefont {Manuel}}, \bibinfo {author}
  {\bibfnamefont {A.~E.}\ \bibnamefont {Trumper}}, \ and\ \bibinfo {author}
  {\bibfnamefont {C.~D.}\ \bibnamefont {Batista}},\ }\href@noop {} {\bibfield
  {journal} {\bibinfo  {journal} {Physical Review B}\ }\textbf {\bibinfo
  {volume} {100}},\ \bibinfo {pages} {104431} (\bibinfo {year}
  {2019})}\BibitemShut {NoStop}%
\bibitem [{\citenamefont {Stiles}\ and\ \citenamefont
  {Zangwill}(2002)}]{prb66.014407}%
  \BibitemOpen
  \bibfield  {author} {\bibinfo {author} {\bibfnamefont {M.~D.}\ \bibnamefont
  {Stiles}}\ and\ \bibinfo {author} {\bibfnamefont {A.}~\bibnamefont
  {Zangwill}},\ }\href@noop {} {\bibfield  {journal} {\bibinfo  {journal}
  {Physical Review B}\ }\textbf {\bibinfo {volume} {66}},\ \bibinfo {pages}
  {014407} (\bibinfo {year} {2002})}\BibitemShut {NoStop}%
\bibitem [{\citenamefont {Garanin}(1996)}]{prb53.11593}%
  \BibitemOpen
  \bibfield  {author} {\bibinfo {author} {\bibfnamefont {D.~A.}\ \bibnamefont
  {Garanin}},\ }\href@noop {} {\bibfield  {journal} {\bibinfo  {journal}
  {Physical Review B}\ }\textbf {\bibinfo {volume} {53}},\ \bibinfo {pages}
  {11593} (\bibinfo {year} {1996})}\BibitemShut {NoStop}%
\bibitem [{\citenamefont {Horwitz}\ and\ \citenamefont
  {Callen}(1961)}]{pr124.1757}%
  \BibitemOpen
  \bibfield  {author} {\bibinfo {author} {\bibfnamefont {G.}~\bibnamefont
  {Horwitz}}\ and\ \bibinfo {author} {\bibfnamefont {H.~B.}\ \bibnamefont
  {Callen}},\ }\href@noop {} {\bibfield  {journal} {\bibinfo  {journal}
  {Physical Review}\ }\textbf {\bibinfo {volume} {124}},\ \bibinfo {pages}
  {1757} (\bibinfo {year} {1961})}\BibitemShut {NoStop}%
\bibitem [{\citenamefont {Stinchcombe}\ \emph {et~al.}(1963)\citenamefont
  {Stinchcombe}, \citenamefont {Horwitz}, \citenamefont {Englert},\ and\
  \citenamefont {Brout}}]{pr130.155}%
  \BibitemOpen
  \bibfield  {author} {\bibinfo {author} {\bibfnamefont {R.}~\bibnamefont
  {Stinchcombe}}, \bibinfo {author} {\bibfnamefont {G.}~\bibnamefont
  {Horwitz}}, \bibinfo {author} {\bibfnamefont {F.}~\bibnamefont {Englert}}, \
  and\ \bibinfo {author} {\bibfnamefont {R.}~\bibnamefont {Brout}},\
  }\href@noop {} {\bibfield  {journal} {\bibinfo  {journal} {Physical Review}\
  }\textbf {\bibinfo {volume} {130}},\ \bibinfo {pages} {155} (\bibinfo {year}
  {1963})}\BibitemShut {NoStop}%
\bibitem [{\citenamefont {Glauber}(1963)}]{pr131.2766}%
  \BibitemOpen
  \bibfield  {author} {\bibinfo {author} {\bibfnamefont {R.~J.}\ \bibnamefont
  {Glauber}},\ }\href@noop {} {\bibfield  {journal} {\bibinfo  {journal}
  {Physical Review}\ }\textbf {\bibinfo {volume} {131}},\ \bibinfo {pages}
  {2766} (\bibinfo {year} {1963})}\BibitemShut {NoStop}%
\bibitem [{\citenamefont {Gerry}\ \emph {et~al.}(2005)\citenamefont {Gerry},
  \citenamefont {Knight},\ and\ \citenamefont {Knight}}]{gerry}%
  \BibitemOpen
  \bibfield  {author} {\bibinfo {author} {\bibfnamefont {C.}~\bibnamefont
  {Gerry}}, \bibinfo {author} {\bibfnamefont {P.}~\bibnamefont {Knight}}, \
  and\ \bibinfo {author} {\bibfnamefont {P.~L.}\ \bibnamefont {Knight}},\
  }\href@noop {} {\emph {\bibinfo {title} {Introductory Quantum Optics}}}\
  (\bibinfo  {publisher} {Cambridge University Press},\ \bibinfo {address}
  {United States of America},\ \bibinfo {year} {2005})\BibitemShut {NoStop}%
\bibitem [{\citenamefont {Rezende}\ and\ \citenamefont
  {Zagury}(1969)}]{pla29.47}%
  \BibitemOpen
  \bibfield  {author} {\bibinfo {author} {\bibfnamefont {S.}~\bibnamefont
  {Rezende}}\ and\ \bibinfo {author} {\bibfnamefont {N.}~\bibnamefont
  {Zagury}},\ }\href@noop {} {\bibfield  {journal} {\bibinfo  {journal}
  {Physics Letters A}\ }\textbf {\bibinfo {volume} {29}},\ \bibinfo {pages}
  {47} (\bibinfo {year} {1969})}\BibitemShut {NoStop}%
\bibitem [{\citenamefont {Zagury}\ and\ \citenamefont
  {Rezende}(1969)}]{pla29.616}%
  \BibitemOpen
  \bibfield  {author} {\bibinfo {author} {\bibfnamefont {N.}~\bibnamefont
  {Zagury}}\ and\ \bibinfo {author} {\bibfnamefont {S.}~\bibnamefont
  {Rezende}},\ }\href@noop {} {\bibfield  {journal} {\bibinfo  {journal}
  {Physics Letters A}\ }\textbf {\bibinfo {volume} {29}},\ \bibinfo {pages}
  {616} (\bibinfo {year} {1969})}\BibitemShut {NoStop}%
\bibitem [{\citenamefont {Zagury}\ and\ \citenamefont
  {Rezende}(1971)}]{prb4.201}%
  \BibitemOpen
  \bibfield  {author} {\bibinfo {author} {\bibfnamefont {N.}~\bibnamefont
  {Zagury}}\ and\ \bibinfo {author} {\bibfnamefont {S.~M.}\ \bibnamefont
  {Rezende}},\ }\href@noop {} {\bibfield  {journal} {\bibinfo  {journal}
  {Physical Review B}\ }\textbf {\bibinfo {volume} {4}},\ \bibinfo {pages}
  {201} (\bibinfo {year} {1971})}\BibitemShut {NoStop}%
\bibitem [{\citenamefont {Zhang}\ \emph {et~al.}(1990)\citenamefont {Zhang},
  \citenamefont {Feng},\ and\ \citenamefont {Gilmore}}]{rmp62.867}%
  \BibitemOpen
  \bibfield  {author} {\bibinfo {author} {\bibfnamefont {W.-M.}\ \bibnamefont
  {Zhang}}, \bibinfo {author} {\bibfnamefont {D.~H.}\ \bibnamefont {Feng}}, \
  and\ \bibinfo {author} {\bibfnamefont {R.}~\bibnamefont {Gilmore}},\
  }\href@noop {} {\bibfield  {journal} {\bibinfo  {journal} {Reviews of Modern
  Physics}\ }\textbf {\bibinfo {volume} {62}},\ \bibinfo {pages} {867}
  (\bibinfo {year} {1990})}\BibitemShut {NoStop}%
\bibitem [{\citenamefont {Pires}\ \emph {et~al.}(1994)\citenamefont {Pires},
  \citenamefont {Pereira},\ and\ \citenamefont {Gouv{\^e}a}}]{prb49.9663}%
  \BibitemOpen
  \bibfield  {author} {\bibinfo {author} {\bibfnamefont {A.~S.~T.}\
  \bibnamefont {Pires}}, \bibinfo {author} {\bibfnamefont {A.~R.}\ \bibnamefont
  {Pereira}}, \ and\ \bibinfo {author} {\bibfnamefont {M.~E.}\ \bibnamefont
  {Gouv{\^e}a}},\ }\href@noop {} {\bibfield  {journal} {\bibinfo  {journal}
  {Physical Review B}\ }\textbf {\bibinfo {volume} {49}},\ \bibinfo {pages}
  {9663} (\bibinfo {year} {1994})}\BibitemShut {NoStop}%
\bibitem [{\citenamefont {Pires}(1995)}]{pla202.309}%
  \BibitemOpen
  \bibfield  {author} {\bibinfo {author} {\bibfnamefont {A.}~\bibnamefont
  {Pires}},\ }\href@noop {} {\bibfield  {journal} {\bibinfo  {journal} {Physics
  Letters A}\ }\textbf {\bibinfo {volume} {202}},\ \bibinfo {pages} {309}
  (\bibinfo {year} {1995})}\BibitemShut {NoStop}%
\bibitem [{\citenamefont {Pereira}\ \emph {et~al.}(1995)\citenamefont
  {Pereira}, \citenamefont {Pires},\ and\ \citenamefont
  {Gouvea}}]{prb51.16413}%
  \BibitemOpen
  \bibfield  {author} {\bibinfo {author} {\bibfnamefont {A.~R.}\ \bibnamefont
  {Pereira}}, \bibinfo {author} {\bibfnamefont {A.~S.~T.}\ \bibnamefont
  {Pires}}, \ and\ \bibinfo {author} {\bibfnamefont {M.~E.}\ \bibnamefont
  {Gouvea}},\ }\href@noop {} {\bibfield  {journal} {\bibinfo  {journal}
  {Physical Review B}\ }\textbf {\bibinfo {volume} {51}},\ \bibinfo {pages}
  {16413} (\bibinfo {year} {1995})}\BibitemShut {NoStop}%
\bibitem [{\citenamefont {Costa}\ \emph {et~al.}(1996)\citenamefont {Costa},
  \citenamefont {Pereira},\ and\ \citenamefont {Pires}}]{prb54.3019}%
  \BibitemOpen
  \bibfield  {author} {\bibinfo {author} {\bibfnamefont {B.~V.}\ \bibnamefont
  {Costa}}, \bibinfo {author} {\bibfnamefont {A.~R.}\ \bibnamefont {Pereira}},
  \ and\ \bibinfo {author} {\bibfnamefont {A.~S.~T.}\ \bibnamefont {Pires}},\
  }\href@noop {} {\bibfield  {journal} {\bibinfo  {journal} {Physical Review
  B}\ }\textbf {\bibinfo {volume} {54}},\ \bibinfo {pages} {3019} (\bibinfo
  {year} {1996})}\BibitemShut {NoStop}%
\bibitem [{\citenamefont {Pires}(1997)}]{ssc104.771}%
  \BibitemOpen
  \bibfield  {author} {\bibinfo {author} {\bibfnamefont {A.~S.~T.}\
  \bibnamefont {Pires}},\ }\href@noop {} {\bibfield  {journal} {\bibinfo
  {journal} {Solid State Communications}\ }\textbf {\bibinfo {volume} {104}},\
  \bibinfo {pages} {771} (\bibinfo {year} {1997})}\BibitemShut {NoStop}%
\bibitem [{\citenamefont {Gouv{\^e}a}\ \emph {et~al.}(1999)\citenamefont
  {Gouv{\^e}a}, \citenamefont {Wysin}, \citenamefont {Leonel}, \citenamefont
  {Pires}, \citenamefont {Kamppeter},\ and\ \citenamefont
  {Mertens}}]{prb59.6229}%
  \BibitemOpen
  \bibfield  {author} {\bibinfo {author} {\bibfnamefont {M.~E.}\ \bibnamefont
  {Gouv{\^e}a}}, \bibinfo {author} {\bibfnamefont {G.~M.}\ \bibnamefont
  {Wysin}}, \bibinfo {author} {\bibfnamefont {S.~A.}\ \bibnamefont {Leonel}},
  \bibinfo {author} {\bibfnamefont {A.~S.~T.}\ \bibnamefont {Pires}}, \bibinfo
  {author} {\bibfnamefont {T.}~\bibnamefont {Kamppeter}}, \ and\ \bibinfo
  {author} {\bibfnamefont {F.~G.}\ \bibnamefont {Mertens}},\ }\href@noop {}
  {\bibfield  {journal} {\bibinfo  {journal} {Physical Review B}\ }\textbf
  {\bibinfo {volume} {59}},\ \bibinfo {pages} {6229} (\bibinfo {year}
  {1999})}\BibitemShut {NoStop}%
\bibitem [{\citenamefont {Menezes}\ \emph {et~al.}(1992)\citenamefont
  {Menezes}, \citenamefont {Gouv{\^e}a},\ and\ \citenamefont
  {Pires}}]{pla166.330}%
  \BibitemOpen
  \bibfield  {author} {\bibinfo {author} {\bibfnamefont {S.}~\bibnamefont
  {Menezes}}, \bibinfo {author} {\bibfnamefont {M.}~\bibnamefont {Gouv{\^e}a}},
  \ and\ \bibinfo {author} {\bibfnamefont {A.~S.~T.}\ \bibnamefont {Pires}},\
  }\href@noop {} {\bibfield  {journal} {\bibinfo  {journal} {Physics Letters
  A}\ }\textbf {\bibinfo {volume} {166}},\ \bibinfo {pages} {330} (\bibinfo
  {year} {1992})}\BibitemShut {NoStop}%
\bibitem [{\citenamefont {Pires}\ and\ \citenamefont
  {Gouvea}(1993)}]{prb48.12698}%
  \BibitemOpen
  \bibfield  {author} {\bibinfo {author} {\bibfnamefont {A.~S.~T.}\
  \bibnamefont {Pires}}\ and\ \bibinfo {author} {\bibfnamefont {M.~E.}\
  \bibnamefont {Gouvea}},\ }\href@noop {} {\bibfield  {journal} {\bibinfo
  {journal} {Physical Review B}\ }\textbf {\bibinfo {volume} {48}},\ \bibinfo
  {pages} {12698} (\bibinfo {year} {1993})}\BibitemShut {NoStop}%
\bibitem [{\citenamefont {Pires}(1994)}]{prb50.9592}%
  \BibitemOpen
  \bibfield  {author} {\bibinfo {author} {\bibfnamefont {A.~S.~T.}\
  \bibnamefont {Pires}},\ }\href@noop {} {\bibfield  {journal} {\bibinfo
  {journal} {Physical Review B}\ }\textbf {\bibinfo {volume} {50}},\ \bibinfo
  {pages} {9592} (\bibinfo {year} {1994})}\BibitemShut {NoStop}%
\bibitem [{\citenamefont {Pires}(1996{\natexlab{a}})}]{ssc100.791}%
  \BibitemOpen
  \bibfield  {author} {\bibinfo {author} {\bibfnamefont {A.}~\bibnamefont
  {Pires}},\ }\href@noop {} {\bibfield  {journal} {\bibinfo  {journal} {Solid
  state communications}\ }\textbf {\bibinfo {volume} {100}},\ \bibinfo {pages}
  {791} (\bibinfo {year} {1996}{\natexlab{a}})}\BibitemShut {NoStop}%
\bibitem [{\citenamefont {Pires}(1996{\natexlab{b}})}]{prb53.235}%
  \BibitemOpen
  \bibfield  {author} {\bibinfo {author} {\bibfnamefont {A.~S.~T.}\
  \bibnamefont {Pires}},\ }\href@noop {} {\bibfield  {journal} {\bibinfo
  {journal} {Physical Review B}\ }\textbf {\bibinfo {volume} {53}},\ \bibinfo
  {pages} {235} (\bibinfo {year} {1996}{\natexlab{b}})}\BibitemShut {NoStop}%
\bibitem [{\citenamefont {Pires}(1996{\natexlab{c}})}]{prb54.6081}%
  \BibitemOpen
  \bibfield  {author} {\bibinfo {author} {\bibfnamefont {A.~S.~T.}\
  \bibnamefont {Pires}},\ }\href@noop {} {\bibfield  {journal} {\bibinfo
  {journal} {Physical Review B}\ }\textbf {\bibinfo {volume} {54}},\ \bibinfo
  {pages} {6081} (\bibinfo {year} {1996}{\natexlab{c}})}\BibitemShut {NoStop}%
\bibitem [{\citenamefont {Pires}(1999)}]{ssc112.705}%
  \BibitemOpen
  \bibfield  {author} {\bibinfo {author} {\bibfnamefont {A.}~\bibnamefont
  {Pires}},\ }\href@noop {} {\bibfield  {journal} {\bibinfo  {journal} {Solid
  state communications}\ }\textbf {\bibinfo {volume} {112}},\ \bibinfo {pages}
  {705} (\bibinfo {year} {1999})}\BibitemShut {NoStop}%
\bibitem [{\citenamefont {Pires}\ and\ \citenamefont
  {Gouv{\^e}a}(2005)}]{epjb2.169}%
  \BibitemOpen
  \bibfield  {author} {\bibinfo {author} {\bibfnamefont {A.~S.~T.}\
  \bibnamefont {Pires}}\ and\ \bibinfo {author} {\bibfnamefont
  {M.}~\bibnamefont {Gouv{\^e}a}},\ }\href@noop {} {\bibfield  {journal}
  {\bibinfo  {journal} {The European Physical Journal B-Condensed Matter and
  Complex Systems}\ }\textbf {\bibinfo {volume} {44}},\ \bibinfo {pages} {169}
  (\bibinfo {year} {2005})}\BibitemShut {NoStop}%
\bibitem [{\citenamefont {Gouv{\^e}a}\ and\ \citenamefont
  {Pires}(2005)}]{pssb.242.2138}%
  \BibitemOpen
  \bibfield  {author} {\bibinfo {author} {\bibfnamefont {M.}~\bibnamefont
  {Gouv{\^e}a}}\ and\ \bibinfo {author} {\bibfnamefont {A.}~\bibnamefont
  {Pires}},\ }\href@noop {} {\bibfield  {journal} {\bibinfo  {journal} {physica
  status solidi (b)}\ }\textbf {\bibinfo {volume} {242}},\ \bibinfo {pages}
  {2138} (\bibinfo {year} {2005})}\BibitemShut {NoStop}%
\bibitem [{\citenamefont {Pires}\ \emph
  {et~al.}(2008{\natexlab{a}})\citenamefont {Pires}, \citenamefont {Costa},\
  and\ \citenamefont {Dias}}]{prb78.212408}%
  \BibitemOpen
  \bibfield  {author} {\bibinfo {author} {\bibfnamefont {A.~S.~T.}\
  \bibnamefont {Pires}}, \bibinfo {author} {\bibfnamefont {B.~V.}\ \bibnamefont
  {Costa}}, \ and\ \bibinfo {author} {\bibfnamefont {R.~A.}\ \bibnamefont
  {Dias}},\ }\href@noop {} {\bibfield  {journal} {\bibinfo  {journal} {Physical
  Review B}\ }\textbf {\bibinfo {volume} {78}},\ \bibinfo {pages} {212408}
  (\bibinfo {year} {2008}{\natexlab{a}})}\BibitemShut {NoStop}%
\bibitem [{\citenamefont {Pires}(2018)}]{jmmm452.315}%
  \BibitemOpen
  \bibfield  {author} {\bibinfo {author} {\bibfnamefont {A.}~\bibnamefont
  {Pires}},\ }\href@noop {} {\bibfield  {journal} {\bibinfo  {journal} {Journal
  of Magnetism and Magnetic Materials}\ }\textbf {\bibinfo {volume} {452}},\
  \bibinfo {pages} {315} (\bibinfo {year} {2018})}\BibitemShut {NoStop}%
\bibitem [{\citenamefont {Pires}(2007)}]{pasma373.387}%
  \BibitemOpen
  \bibfield  {author} {\bibinfo {author} {\bibfnamefont {A.}~\bibnamefont
  {Pires}},\ }\href@noop {} {\bibfield  {journal} {\bibinfo  {journal} {Physica
  A: Statistical Mechanics and its Applications}\ }\textbf {\bibinfo {volume}
  {373}},\ \bibinfo {pages} {387} (\bibinfo {year} {2007})}\BibitemShut
  {NoStop}%
\bibitem [{\citenamefont {Pires}\ \emph
  {et~al.}(2008{\natexlab{b}})\citenamefont {Pires}, \citenamefont {Lima},\
  and\ \citenamefont {Gouvea}}]{jpcm20.015208}%
  \BibitemOpen
  \bibfield  {author} {\bibinfo {author} {\bibfnamefont {A.}~\bibnamefont
  {Pires}}, \bibinfo {author} {\bibfnamefont {L.}~\bibnamefont {Lima}}, \ and\
  \bibinfo {author} {\bibfnamefont {M.}~\bibnamefont {Gouvea}},\ }\href@noop {}
  {\bibfield  {journal} {\bibinfo  {journal} {Journal of Physics: Condensed
  Matter}\ }\textbf {\bibinfo {volume} {20}},\ \bibinfo {pages} {015208}
  (\bibinfo {year} {2008}{\natexlab{b}})}\BibitemShut {NoStop}%
\bibitem [{\citenamefont {Pires}\ and\ \citenamefont
  {Gouvea}(2009)}]{pasma388.21}%
  \BibitemOpen
  \bibfield  {author} {\bibinfo {author} {\bibfnamefont {A.}~\bibnamefont
  {Pires}}\ and\ \bibinfo {author} {\bibfnamefont {M.}~\bibnamefont {Gouvea}},\
  }\href@noop {} {\bibfield  {journal} {\bibinfo  {journal} {Physica A:
  Statistical Mechanics and its Applications}\ }\textbf {\bibinfo {volume}
  {388}},\ \bibinfo {pages} {21} (\bibinfo {year} {2009})}\BibitemShut
  {NoStop}%
\bibitem [{\citenamefont {Pires}\ and\ \citenamefont
  {Costa}(2009)}]{pasma388.3779}%
  \BibitemOpen
  \bibfield  {author} {\bibinfo {author} {\bibfnamefont {A.}~\bibnamefont
  {Pires}}\ and\ \bibinfo {author} {\bibfnamefont {B.}~\bibnamefont {Costa}},\
  }\href@noop {} {\bibfield  {journal} {\bibinfo  {journal} {Physica A:
  Statistical Mechanics and its Applications}\ }\textbf {\bibinfo {volume}
  {388}},\ \bibinfo {pages} {3779} (\bibinfo {year} {2009})}\BibitemShut
  {NoStop}%
\bibitem [{\citenamefont {Moura}\ \emph {et~al.}(2014)\citenamefont {Moura},
  \citenamefont {Pires},\ and\ \citenamefont {Pereira}}]{jmmm357.45}%
  \BibitemOpen
  \bibfield  {author} {\bibinfo {author} {\bibfnamefont {A.~R.}\ \bibnamefont
  {Moura}}, \bibinfo {author} {\bibfnamefont {A.~S.}\ \bibnamefont {Pires}}, \
  and\ \bibinfo {author} {\bibfnamefont {A.~R.}\ \bibnamefont {Pereira}},\
  }\href@noop {} {\bibfield  {journal} {\bibinfo  {journal} {Journal of
  magnetism and magnetic materials}\ }\textbf {\bibinfo {volume} {357}},\
  \bibinfo {pages} {45} (\bibinfo {year} {2014})}\BibitemShut {NoStop}%
\bibitem [{\citenamefont {Moura}\ and\ \citenamefont
  {Lopes}(2019)}]{jmmm472.1}%
  \BibitemOpen
  \bibfield  {author} {\bibinfo {author} {\bibfnamefont {A.~R.}\ \bibnamefont
  {Moura}}\ and\ \bibinfo {author} {\bibfnamefont {R.~J.~C.}\ \bibnamefont
  {Lopes}},\ }\href@noop {} {\bibfield  {journal} {\bibinfo  {journal} {Journal
  of Magnetism and Magnetic Materials}\ }\textbf {\bibinfo {volume} {472}},\
  \bibinfo {pages} {1} (\bibinfo {year} {2019})}\BibitemShut {NoStop}%
\bibitem [{\citenamefont {Tserkovnyak}\ \emph {et~al.}(2005)\citenamefont
  {Tserkovnyak}, \citenamefont {Brataas}, \citenamefont {Bauer},\ and\
  \citenamefont {Halperin}}]{rmp77.1375}%
  \BibitemOpen
  \bibfield  {author} {\bibinfo {author} {\bibfnamefont {Y.}~\bibnamefont
  {Tserkovnyak}}, \bibinfo {author} {\bibfnamefont {A.}~\bibnamefont
  {Brataas}}, \bibinfo {author} {\bibfnamefont {G.~E.~W.}\ \bibnamefont
  {Bauer}}, \ and\ \bibinfo {author} {\bibfnamefont {B.~I.}\ \bibnamefont
  {Halperin}},\ }\href@noop {} {\bibfield  {journal} {\bibinfo  {journal}
  {Reviews of Modern Physics}\ }\textbf {\bibinfo {volume} {77}},\ \bibinfo
  {pages} {1375} (\bibinfo {year} {2005})}\BibitemShut {NoStop}%
\bibitem [{\citenamefont {Mahan}(2013)}]{mahan}%
  \BibitemOpen
  \bibfield  {author} {\bibinfo {author} {\bibfnamefont {G.~D.}\ \bibnamefont
  {Mahan}},\ }\href@noop {} {\emph {\bibinfo {title} {Many-Particle Physics}}}\
  (\bibinfo  {publisher} {Springer Science \& Business Media},\ \bibinfo
  {address} {United States of America},\ \bibinfo {year} {2013})\BibitemShut
  {NoStop}%
\bibitem [{\citenamefont {Kondo}(1964)}]{ptp32.37}%
  \BibitemOpen
  \bibfield  {author} {\bibinfo {author} {\bibfnamefont {J.}~\bibnamefont
  {Kondo}},\ }\href@noop {} {\bibfield  {journal} {\bibinfo  {journal}
  {Progress of theoretical physics}\ }\textbf {\bibinfo {volume} {32}},\
  \bibinfo {pages} {37} (\bibinfo {year} {1964})}\BibitemShut {NoStop}%
\bibitem [{\citenamefont {Villain}(1974)}]{jp35.27}%
  \BibitemOpen
  \bibfield  {author} {\bibinfo {author} {\bibfnamefont {J.}~\bibnamefont
  {Villain}},\ }\href@noop {} {\bibfield  {journal} {\bibinfo  {journal}
  {Journal de Physique}\ }\textbf {\bibinfo {volume} {35}},\ \bibinfo {pages}
  {27} (\bibinfo {year} {1974})}\BibitemShut {NoStop}%
\bibitem [{\citenamefont {Mann}\ and\ \citenamefont
  {Revzen}(1989)}]{pla134.273}%
  \BibitemOpen
  \bibfield  {author} {\bibinfo {author} {\bibfnamefont {A.}~\bibnamefont
  {Mann}}\ and\ \bibinfo {author} {\bibfnamefont {M.}~\bibnamefont {Revzen}},\
  }\href@noop {} {\bibfield  {journal} {\bibinfo  {journal} {Physics Letters
  A}\ }\textbf {\bibinfo {volume} {134}},\ \bibinfo {pages} {273} (\bibinfo
  {year} {1989})}\BibitemShut {NoStop}%
\bibitem [{\citenamefont {Oz-Vogt}\ \emph {et~al.}(1991)\citenamefont
  {Oz-Vogt}, \citenamefont {Mann},\ and\ \citenamefont {Revzen}}]{jmo38.2339}%
  \BibitemOpen
  \bibfield  {author} {\bibinfo {author} {\bibfnamefont {J.}~\bibnamefont
  {Oz-Vogt}}, \bibinfo {author} {\bibfnamefont {A.}~\bibnamefont {Mann}}, \
  and\ \bibinfo {author} {\bibfnamefont {M.}~\bibnamefont {Revzen}},\
  }\href@noop {} {\bibfield  {journal} {\bibinfo  {journal} {Journal of Modern
  Optics}\ }\textbf {\bibinfo {volume} {38}},\ \bibinfo {pages} {2339}
  (\bibinfo {year} {1991})}\BibitemShut {NoStop}%
\bibitem [{\citenamefont {Bloembergen}(1950)}]{pr78.572}%
  \BibitemOpen
  \bibfield  {author} {\bibinfo {author} {\bibfnamefont {N.}~\bibnamefont
  {Bloembergen}},\ }\href@noop {} {\bibfield  {journal} {\bibinfo  {journal}
  {Physical Review}\ }\textbf {\bibinfo {volume} {78}},\ \bibinfo {pages} {572}
  (\bibinfo {year} {1950})}\BibitemShut {NoStop}%
\bibitem [{\citenamefont {Bloembergen}\ and\ \citenamefont
  {Wang}(1954)}]{pr93.72}%
  \BibitemOpen
  \bibfield  {author} {\bibinfo {author} {\bibfnamefont {N.}~\bibnamefont
  {Bloembergen}}\ and\ \bibinfo {author} {\bibfnamefont {S.}~\bibnamefont
  {Wang}},\ }\href@noop {} {\bibfield  {journal} {\bibinfo  {journal} {Physical
  Review}\ }\textbf {\bibinfo {volume} {93}},\ \bibinfo {pages} {72} (\bibinfo
  {year} {1954})}\BibitemShut {NoStop}%
\bibitem [{\citenamefont {Yal{\c{c}}{\i}n}(2013)}]{yalccin}%
  \BibitemOpen
  \bibfield  {author} {\bibinfo {author} {\bibfnamefont {O.}~\bibnamefont
  {Yal{\c{c}}{\i}n}},\ }\href@noop {} {\emph {\bibinfo {title} {Ferromagnetic
  resonance: theory and applications}}}\ (\bibinfo  {publisher} {BoD--Books on
  Demand},\ \bibinfo {address} {Croatia},\ \bibinfo {year} {2013})\BibitemShut
  {NoStop}%
\bibitem [{\citenamefont {Rezende}\ \emph {et~al.}(2013)\citenamefont
  {Rezende}, \citenamefont {Rodr{\'\i}guez-Su{\'a}rez}, \citenamefont {Soares},
  \citenamefont {Vilela-Le{\~a}o}, \citenamefont {Ley~Dom{\'\i}nguez},\ and\
  \citenamefont {Azevedo}}]{apl102.012402}%
  \BibitemOpen
  \bibfield  {author} {\bibinfo {author} {\bibfnamefont {S.}~\bibnamefont
  {Rezende}}, \bibinfo {author} {\bibfnamefont {R.}~\bibnamefont
  {Rodr{\'\i}guez-Su{\'a}rez}}, \bibinfo {author} {\bibfnamefont
  {M.}~\bibnamefont {Soares}}, \bibinfo {author} {\bibfnamefont
  {L.}~\bibnamefont {Vilela-Le{\~a}o}}, \bibinfo {author} {\bibfnamefont
  {D.}~\bibnamefont {Ley~Dom{\'\i}nguez}}, \ and\ \bibinfo {author}
  {\bibfnamefont {A.}~\bibnamefont {Azevedo}},\ }\href@noop {} {\bibfield
  {journal} {\bibinfo  {journal} {Applied Physics Letters}\ }\textbf {\bibinfo
  {volume} {102}},\ \bibinfo {pages} {012402} (\bibinfo {year}
  {2013})}\BibitemShut {NoStop}%
\bibitem [{\citenamefont {Heinrich}\ \emph {et~al.}(2011)\citenamefont
  {Heinrich}, \citenamefont {Burrowes}, \citenamefont {Montoya}, \citenamefont
  {Kardasz}, \citenamefont {Girt}, \citenamefont {Song}, \citenamefont {Sun},\
  and\ \citenamefont {Wu}}]{prl107.066604}%
  \BibitemOpen
  \bibfield  {author} {\bibinfo {author} {\bibfnamefont {B.}~\bibnamefont
  {Heinrich}}, \bibinfo {author} {\bibfnamefont {C.}~\bibnamefont {Burrowes}},
  \bibinfo {author} {\bibfnamefont {E.}~\bibnamefont {Montoya}}, \bibinfo
  {author} {\bibfnamefont {B.}~\bibnamefont {Kardasz}}, \bibinfo {author}
  {\bibfnamefont {E.}~\bibnamefont {Girt}}, \bibinfo {author} {\bibfnamefont
  {Y.-Y.}\ \bibnamefont {Song}}, \bibinfo {author} {\bibfnamefont
  {Y.}~\bibnamefont {Sun}}, \ and\ \bibinfo {author} {\bibfnamefont
  {M.}~\bibnamefont {Wu}},\ }\href@noop {} {\bibfield  {journal} {\bibinfo
  {journal} {Physical Review Letters}\ }\textbf {\bibinfo {volume} {107}},\
  \bibinfo {pages} {066604} (\bibinfo {year} {2011})}\BibitemShut {NoStop}%
\bibitem [{\citenamefont {Takahashi}\ \emph {et~al.}(2012)\citenamefont
  {Takahashi}, \citenamefont {Iguchi}, \citenamefont {Ando}, \citenamefont
  {Nakayama}, \citenamefont {Yoshino},\ and\ \citenamefont
  {Saitoh}}]{jap111.07c307}%
  \BibitemOpen
  \bibfield  {author} {\bibinfo {author} {\bibfnamefont {R.}~\bibnamefont
  {Takahashi}}, \bibinfo {author} {\bibfnamefont {R.}~\bibnamefont {Iguchi}},
  \bibinfo {author} {\bibfnamefont {K.}~\bibnamefont {Ando}}, \bibinfo {author}
  {\bibfnamefont {H.}~\bibnamefont {Nakayama}}, \bibinfo {author}
  {\bibfnamefont {T.}~\bibnamefont {Yoshino}}, \ and\ \bibinfo {author}
  {\bibfnamefont {E.}~\bibnamefont {Saitoh}},\ }\href@noop {} {\bibfield
  {journal} {\bibinfo  {journal} {Journal of Applied Physics}\ }\textbf
  {\bibinfo {volume} {111}},\ \bibinfo {pages} {07C307} (\bibinfo {year}
  {2012})}\BibitemShut {NoStop}%
\bibitem [{\citenamefont {Du}\ \emph {et~al.}(2014)\citenamefont {Du},
  \citenamefont {Wang}, \citenamefont {Yang},\ and\ \citenamefont
  {Hammel}}]{pra1.044004}%
  \BibitemOpen
  \bibfield  {author} {\bibinfo {author} {\bibfnamefont {C.}~\bibnamefont
  {Du}}, \bibinfo {author} {\bibfnamefont {H.}~\bibnamefont {Wang}}, \bibinfo
  {author} {\bibfnamefont {F.}~\bibnamefont {Yang}}, \ and\ \bibinfo {author}
  {\bibfnamefont {P.~C.}\ \bibnamefont {Hammel}},\ }\href@noop {} {\bibfield
  {journal} {\bibinfo  {journal} {Physical Review Applied}\ }\textbf {\bibinfo
  {volume} {1}},\ \bibinfo {pages} {044004} (\bibinfo {year}
  {2014})}\BibitemShut {NoStop}%
\end{thebibliography}%
\end{document}